\documentclass[12pt]{JHEP3}

\usepackage{amssymb}
\usepackage{graphicx}
\usepackage{amsmath}
\usepackage{amsfonts}

\newcommand{\beq}{\begin{equation}}
\newcommand{\eeq}{\end{equation}}
\newcommand{\ben}{\begin{enumerate}}
\newcommand{\een}{\end{enumerate}}
\newcommand{\bit}{\begin{itemize}}
\newcommand{\eit}{\end{itemize}}
\newcommand{\beqray}{\begin{eqnarray}}
\newcommand{\eeqray}{\end{eqnarray}}

\newcommand{\dd}{\mathrm{d}}

\newcommand{\hinvmpc}{h^{-1}\mathrm{Mpc}}

\newcommand{\hmpcinv}{h\mathrm{Mpc}^{-1}}

\newcommand{\Omegam}{\Omega_{\mathrm{M}}}
\newcommand{\Omegab}{\Omega_{\mathrm{b}}}

\newcommand{\ns}{n_{\mathrm{s}}}
\newcommand{\cmbnorm}{\Delta_{\mathcal{R}}^{2}}

\newcommand{\da}{\mathrm{D_A}}
\newcommand{\daz}{\mathrm{D_A}(z)}

\newcommand{\dasqz}{\mathrm{D}_{\mathrm{A}}^{2}(z)}

\newcommand{\wzero}{w_{0}}
\newcommand{\wa}{w_{\mathrm{a}}}

\newcommand{\wpiv}{w_{\mathrm{piv}}}

\newcommand{\pgg}{P_{\mathrm{gg}}}
\newcommand{\pkg}{P_{\kappa\mathrm{g}}}

\newcommand{\kappai}{\kappa_{\mathrm{i}}}
\newcommand{\kappaj}{\kappa_{\mathrm{j}}}

\newcommand{\gi}{g_{\mathrm{i}}}
\newcommand{\gj}{g_{\mathrm{j}}}

\newcommand{\pkikj}{P_{\kappai\kappaj}}
\newcommand{\pgigj}{P_{\gi\gj}}
\newcommand{\pkigj}{P_{\kappai\gj}}

\newcommand{\biz}{\mathrm{b}_{\mathrm{i}}(z)}

\newcommand{\ssubi}{\mathrm{s}_{\mathrm{i}}}
\newcommand{\ssubj}{\mathrm{s}_{\mathrm{j}}}

\newcommand{\xsubi}{\mathrm{x}_{\mathrm{i}}}
\newcommand{\xsubj}{\mathrm{x}_{\mathrm{j}}}

\newcommand{\xsubm}{\mathrm{x}_{\mathrm{m}}}
\newcommand{\xsubn}{\mathrm{x}_{\mathrm{n}}}

\newcommand{\wiz}{\mathrm{W}_{i}(z)}
\newcommand{\wjz}{\mathrm{W}_{j}(z)}

\newcommand{\niz}{n_\mathrm{{i}}(z)}
\newcommand{\phzpdf}{P(z^{ph}|z)}
\newcommand{\zphot}{z^{\mathrm{ph}}}
\newcommand{\na}{N_{\mathrm{A}}}
\newcommand{\zbias}{z_{\mathrm{bias}}}

\newcommand{\nspec}{\mathrm{N_{spec}}}
\newcommand{\nspeci}{\mathrm{N_{spec}^{i}}}

\newcommand{\czero}{c_{0}}

\newcommand{\pmatk}{\mathrm{P}_{\delta}(k)}

\newcommand{\pmatkz}{\mathrm{P}_{\delta}(k,z)}

\newcommand{\dpmati}{\delta \mathrm{ln}(P_{\mathrm{i}})}

\newcommand{\kreq}{k_{\mathrm{req}}^{\mathrm{max}}}
\newcommand{\ith}{i^{\mathrm{th}}}

\newcommand{\fsky}{f_{\mathrm{sky}}}
\newcommand{\ellmax}{\ell_{\mathrm{max}}}

\newcommand{\fishinvab}{\mathcal{F}^{-1}_{\alpha\beta}}
\newcommand{\fishinvaa}{\mathcal{F}^{-1}_{\alpha\alpha}}
\newcommand{\fishinvbb}{\mathcal{F}^{-1}_{\beta\beta}}

\newcommand{\qij}{\mathcal{Q}_{\mathrm{ij}}}
\newcommand{\qab}{\mathcal{Q}_{\alpha\beta}}
\newcommand{\pari}{p_{\mathrm{i}}}
\newcommand{\parj}{p_{\mathrm{j}}}
\newcommand{\para}{p_{\alpha}}
\newcommand{\parb}{p_{\beta}}

\title{
General Requirements on Matter Power Spectrum Predictions for Cosmology with Weak Lensing Tomography
}

\author{Andrew P. Hearin and Andrew R. Zentner\\
Pittsburgh Particle physics Astrophysics and Cosmology Center (PITT PACC)\\ 
Department of Physics and Astronomy, University of Pittsburgh,\\ 
Pittsburgh, PA 15260 USA
}
\author{Zhaoming Ma\\
Brookhaven National Laboratory, Upton, NY 11973
}

\abstract{
Forthcoming projects such as DES, LSST, WFIRST, and Euclid aim to measure weak lensing shear correlations with unprecedented precision, constraining the dark energy equation of state at the percent level. Reliance on photometrically-determined redshifts constitutes a major source of uncertainty for these surveys. Additionally, interpreting the weak lensing signal requires a detailed understanding of the nonlinear physics of gravitational collapse. We present a new analysis of the stringent calibration requirements for weak lensing analyses of future imaging surveys that addresses both photo-z uncertainty and errors in the calibration of the matter power spectrum. We find that when photo-z uncertainty is taken into account the requirements on the level of precision in the prediction for the matter power spectrum are more stringent than previously thought. Including degree-scale galaxy clustering statistics in a joint analysis with weak lensing not only strengthens the survey's constraining power by $\sim20\%,$ but can also have a profound impact on the calibration demands, decreasing the degradation in dark energy constraints with matter power spectrum uncertainty by a factor of $2-5.$ Similarly, using galaxy clustering information significantly relaxes the demands on photo-z calibration. We compare these calibration requirements to the contemporary state-of-the-art in photometric redshift estimation and predictions of the power spectrum and suggest strategies to utilize forthcoming data optimally.
}


\begin{document}

\section{Introduction}

Weak gravitational lensing of galaxies by large-scale structure is a potentially powerful cosmological probe 
\cite{hoekstra_etal02,pen_etal03,jarvis_etal03,van_waerbeke_etal05,jarvis_etal06,semboloni_etal06,kitching_etal07,benjamin_etal07,dore_etal07,fu_etal08,weinberg_etal12}.  
Forthcoming imaging surveys such as the Dark Energy Survey (DES), 
the survey to be conducted by the Large Synoptic Survey Telescope (LSST), the survey of the 
European Space Agency's Euclid satellite \cite{euclid_redbook}, and the proposed Wide Field Infra-Red Survey Telescope (WFIRST) 
expect to exploit measurements of weak gravitational lensing of distant source galaxies to constrain the properties 
of the dark energy \cite{hu_tegmark99,hu99,huterer02,heavens03,refregier03,refregier_etal04,
song_knox04,takada_jain04,takada_white04,dodelson_zhang05,ishak05,albrecht_etal06,
zhan06,munshi_etal08,hoekstra_jain08,zentner_etal08,zhao_etal09,hearin_etal10}.  
Over the last several years, it has been recognized that the limited precision with which the 
matter power spectrum can be predicted may be one of several important, systematic errors 
that will need to be controlled in order to realize this goal (e.g., 
Refs.~\cite{white04,zhan_knox04,huterer_takada05,jing_etal06,eifler11,rudd_etal08,zentner_etal08,guillet_etal10,casarini_etal11,semboloni_etal11}).  

A large part of this uncertainty is due to the effects of baryons on lensing power spectra, which were largely neglected 
in much of the early literature on cosmological weak lensing.  Several groups, including our own, have begun numerical 
simulation programs designed to address this issue with large-scale numerical simulations 
(e.g., Ref.~\cite{semboloni_etal11,heitmann_etal08,heitmann_etal08b,heitmann_etal09,bhattacharya_etal11,casarini_etal11b}).  With the notable exception 
of Ref.~\cite{huterer_takada05}, there have not been detailed studies of the precision with which the matter power spectrum 
must be predicted before it becomes a relatively small contributor to the error budget.  
In the run-up to large, computationally-intensive and human resource-intensive simulation campaigns, 
we have studied the required precision with which the matter power spectrum must be predicted in order to realize anticipated dark energy 
constraints from cosmic weak lensing tomography.  We present our results as a set of general guidelines 
on the systematic errors on the matter power spectrum as a function of scale and redshift.  

Another potentially dominant source of error for dark energy parameter estimators arises from the necessity of using approximate 
redshifts determined from photometric data rather than spectroscopic redshifts 
\cite{ma_etal06,huterer_etal06,lima_hu07,kitching_etal08,ma_bernstein08,newman08,sun_etal09,
zentner_bhattacharya09,bernstein_huterer09,zhang_etal09}.  
Photometric redshifts are significantly less precise than spectroscopic redshifts, and can 
exhibit large biases. Interestingly, we find that the precision with which the power spectrum must be predicted 
is very sensitive to the precision of photometric redshift determinations and vice versa, a result hinted at 
in the Appendix of Ref.~\cite{hearin_etal10}.  As part of our analysis, we model photometric 
redshift uncertainty and show how the precision with which the power spectrum must be predicted 
varies with photometric redshift errors, and conversely. 

Briefly, we find that if prior information on the photometric redshift distribution is weak, then dark energy constraints degrade $2-3$ times more rapidly with uncertainty in $\pmatk$ than if the photo-z distribution is characterized with high precision. Thus we find that when photo-z uncertainty is taken into account the calibration requirements on the theoretical prediction for the matter power spectrum are more stringent than previously thought. The complementarity of galaxy clustering statistics with weak lensing, well-studied in other contexts (see, for example, Refs.~\cite{zhan06},~\cite{zhan_etal08},~\cite{shapiro_dodelson07},~\cite{joudaki_kaplinghat11}, and~\cite{zhao_etal11}), has an ameliorating effect on power spectrum misestimations. Even when restricted to degree-scales, including galaxy correlation information can mitigate dark energy systematics induced by errors in the prediction for $\pmatk$ by up to $50\%;$ alternatively, neglecting galaxy clustering statistics can cause the statistical constraints on dark energy parameters to degrade $2-5$ times more rapidly with uncertainty in either $\pmatk$ or the photo-z distribution. 

This manuscript is organized as follows. In~\S\ref{section:methods} we describe how we model uncertainty in photometric redshifts 
and in $\pmatk$ as well as our methods for estimating statistical and systematic errors. We present our results in~\S\ref{section:results} and discuss their implications in~\S\ref{section:discussion}. We conclude in~\S\ref{section:conclusions} with a summary of our primary results.

%----------------------------------------------------------------------------------------------------------------------------------------
%----------------------------------------------------------------------------------------------------------------------------------------
\section{Methods}
\label{section:methods}

\subsection{The Matter Power Spectrum}
\label{subsection:nlcompare}

A significant amount of the constraining power of weak lensing surveys will come from scales on 
which the nonlinear effects of gravitational collapse cannot be neglected \cite{zentner_etal08,huterer_white05,zhan_knox06}.  
It is possible to excise data on relatively small scales, but such an excision significantly degrades dark 
energy constraints from cosmological weak lensing \cite{zentner_etal08,huterer_white05}.  Modeling nonlinear 
structure formation will be essential in order for forthcoming galaxy imaging surveys to realize their potential 
for constraining dark energy and modified gravity \cite{huterer_takada05,hearin_zentner09}.  As this modeling is uncertain and 
can be an important source of error, we study the errors induced on dark energy parameter estimators by 
uncertainty in the underlying matter power spectrum and we quantify the relative importance of theoretical 
power spectrum errors as a function of wavenumber.  Our results may 
serve as a guideline for computational programs aimed at predicting accurate and precise matter power spectra 
for the purpose of comparing with weak lensing data.

\begin{figure*}[t!]
\centering
\includegraphics[width=14.0cm]{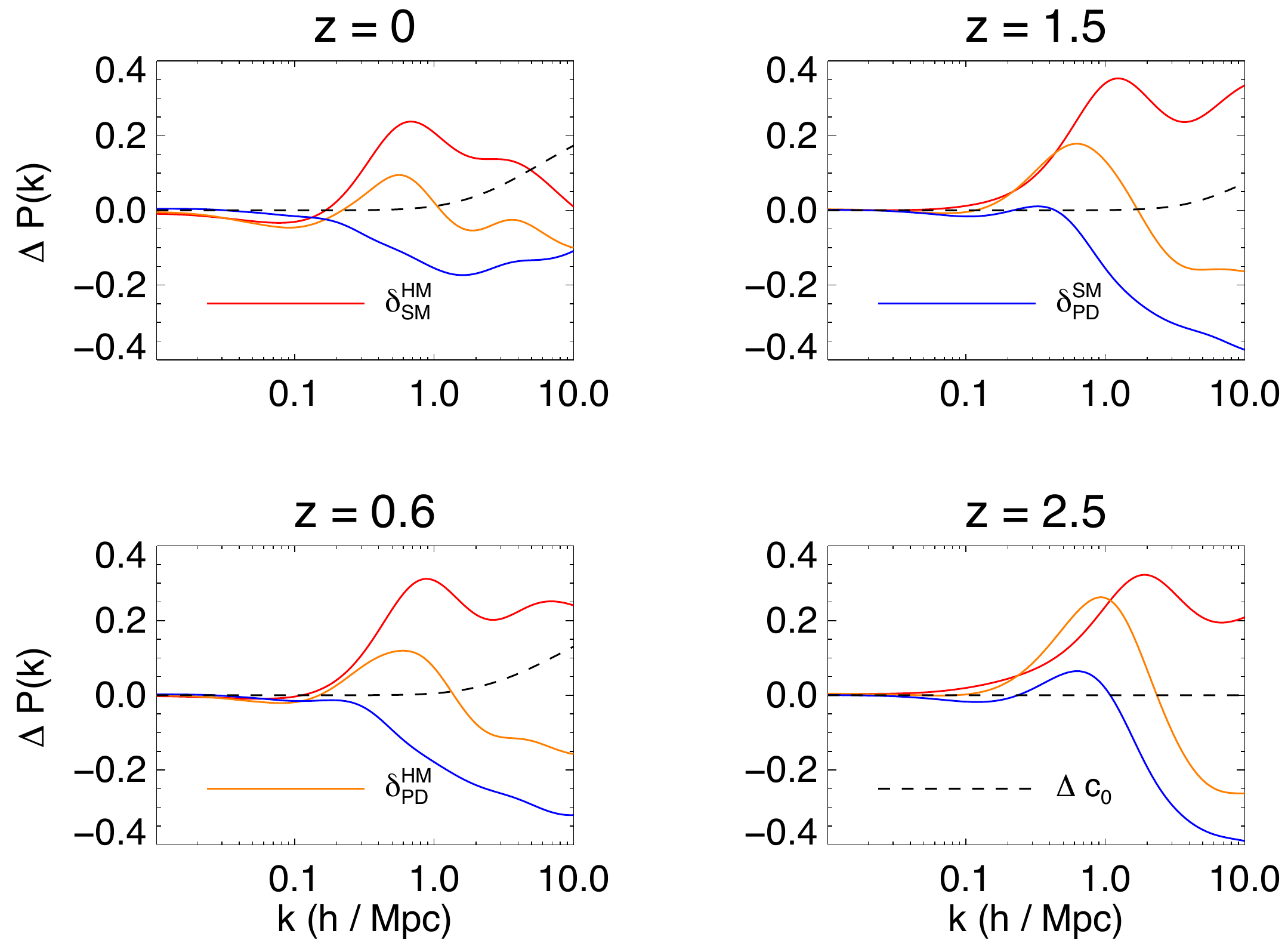}
\caption{
Fractional difference in $\pmatk$ as predicted by the three different models of nonlinear evolution.
The superscript indicates the fiducial model, the subscript the model inducing the fractional change.  
For example, the curve labelled by $\delta^{\mathrm{HM}}_{\mathrm{SM}}$ represents the fractional 
change to the $\pmatk$ induced by using Smith et al. to compute nonlinear power rather than the fiducial halo model; 
thus when this curve is positive the Smith et al. prediction for $\pmatk$ exceeds that of the halo model.
The dashed curve labeled $\Delta c_{0}$ shows the fractional change 
to $\pmatk$ induced by systematically misestimating the mean concentration of dark matter halos by $20\%.$ 
}
\label{fig:deltak_wrongnl}
\end{figure*}

In the current and past literature, the three most commonly-used techniques employed for predicting the 
matter power spectrum in the mildly nonlinear regime are the fitting formula of Peacock \& Dodds \cite{peacock_dodds96}, 
the Halo Model (see Ref.~\cite{cooray_sheth02} for a review, as 
well as the many references therein), and the fitting formula of Smith et al.~\cite{smith_etal03}.  
As a rough look at the contemporary level of uncertainty in predictions for the matter power spectrum $\pmatk$, and to 
set the stage for what is to come, we have plotted in Fig.~\ref{fig:deltak_wrongnl} the fractional difference in $\pmatk$ at several 
different redshifts predicted by these three nonlinear evolution models.  The differences in the predictions made by these methods 
become significant on scales ($k \gtrsim0.2 \hmpcinv$), which coincides with the scales at which the constraining power of weak lensing begins to peak. When the baryonic physics of galaxy formation is taken into account (for example, as in Refs.~\cite{rudd_etal08,semboloni_etal11,guillet_etal10}), among other possible effects the matter distribution within halos is known to change relative to N-body (dark matter-only) simulations. To illustrate the effect such a rearrangement may have on the matter power spectrum, in Fig.~\ref{fig:deltak_wrongnl} we additionally plot the fractional difference in $\pmatk$ that is induced when the Halo Model correctly predicts the nonlinear evolution but the mean concentration of dark matter halos (quantified by the parameter $c_{0}$ in Eq.~\ref{eq:haloprofile}) is misestimated by $20\%.$ 

%---------------------------------------------------------------------------------------------------
%-----------------------------------------------------------------------------------------------------

In this paper, we attempt to remain relatively agnostic about the types of errors that may be realized 
in predictions of the matter power spectrum at the percent and sub-percent levels, because such accurate 
predictions have not yet been made.  We study two classes of models for uncertainty in the matter power spectrum.  
Our first model is motivated by simulation results suggesting that one large source of systematic error in nonlinear 
power spectrum predictions may arise from a systematic error in predictions for the internal structures of dark matter 
halos.  Significant rearrangement of dark matter may be a result of baryonic processes during galaxy formation, 
for example \cite{jing_etal06,rudd_etal08,guillet_etal10}.  So, we suppose that halo abundances and halo clustering 
are well-known so that the halo model accurately predicts the gross shape of $\pmatk,$ but that the distribution of matter 
within halos is uncertain.  For our purposes, the average halo density profile may be taken as the Navarro-Frenk-White (NFW) 
density profile \cite{navarro_etal97}, 

%----------------------
\beq
\label{eq:haloprofile}
\rho(r) \propto \left( c\,\frac{r}{R_{\mathrm{200m}}} \right)^{-1}\, \left(1+c\,\frac{r}{R_{\mathrm{200m}}} \right)^{-2}.
\eeq
The parameter $c$ describes the concentration of the mass distribution towards the halo center; 
the mass contained within the radius $R_{\mathrm{200m}}$ defines the halo mass. 
The average halo concentration varies with halo mass and evolves in redshift; we model these dependences as a power laws, 

%-----------------
\beq
\label{eq:cmz}
c(m,z)=c_{0}[m/m_{*,0}]^{\alpha}(1+z)^{\beta},
\eeq
%---------------
%
where $m_{*,0}=2.2\times10^{12}$ $M_{\odot}/h$ is the typical mass of a halo that is just starting to collapse at $z=0.$ We allow the parameters  $\alpha, \beta, c_0$ to vary about their fiducial values of $c_0=10.7,$ $\alpha=-0.11,$ and $\beta=-1,$ 
so that in this model uncertainty in $\pmatk$ stems exclusively from uncertainty in the distribution of mass 
within gravitationally-collapsed, self-bound objects. Our fiducial values for $\alpha$ and $\beta$ are chosen to match results from the Millennium Simulation \cite{neto_etal07}, but our fiducial value of $c_0$ is roughly a factor of two larger than the mean concentration of Millennium Simulation redshift-zero halos. This larger value is motivated by the results from Ref.~\cite{rudd_etal08} showing that baryonic physics typically produces a significant enhancement to the concentrations of dark matter halos. 
We thus set our fiducial $c_0$ according to Ref.~\cite{rudd_etal08} rather than the Millennium results.
 
Our second class of model for uncertainty in the matter power spectrum $\pmatkz$, is significantly more general.  
After choosing a technique for predicting a {\em fiducial} matter power spectrum on scales of interest, we parameterize 
uncertainty in $\pmatk$ as follows.  The range of scales between $k=0.01\hmpcinv$ and $k=10 \hmpcinv$ are binned evenly 
in $\mathrm{log}(k),$ and uncertainty in the matter power spectrum on a scale $k$ in the $\ith$ bin is parameterized via an 
additional parameter $\dpmati$, such that 
$$\pmatkz\rightarrow\pmatkz\left(1+\dpmati\right).$$ 
The parameters $\dpmati$ are then allowed to freely vary about their fiducial values of zero.
We discuss our choice for the number of $\dpmati$ parameters in ~\S\ref{sub:forecasting}. 
This scheme for studying $\pmatk$ uncertainty was first introduced in the context of 
weak lensing in \cite{huterer_takada05}. 
%-----------------------------------------------------------------------------------------------------------------
\subsection{Photo-z Uncertainty}
\label{sub:photoz}

As uncertainty in photometric redshifts is likely to be one of the chief contributions to the error budget in future lensing measurements of 
dark energy, forecasts of dark energy parameter constraints need to include photo-z uncertainty to realistically estimate the constraining power of future surveys. While it is the primary goal of this manuscript to study the influence of uncertainty in the matter power spectrum, our previous work has shown that photo-z calibration requirements can depend sensitively on the fiducial model for the nonlinear evolution of $\pmatk.$ In the Appendix of Ref.~\cite{hearin_etal10}, we showed that as prior information on the photometric redshift distribution decreases, constraints on $\wzero$ degrade $\sim5$ times faster if the Peacock \& Dodds fitting formula is used to model nonlinear power   rather than the fitting formula of Smith et al.~(halo model results are intermediary between these two). The primary reason for this difference stems from Smith et al.~predicting more small-scale power relative to the other two models, a trend that is apparent in Fig.~\ref{fig:deltak_wrongnl}. These results demonstrate that there is a nontrivial interplay between the nonlinear evolution of $\pmatk$ and the redshift distribution of the sources used to measure the weak lensing signal. This suggests that uncertainty in photo-z's and $\pmatk$ need to be treated simultaneously in order to accurately predict the calibration requirements for the matter power spectrum. 

We model the underlying redshift distribution as $n(z)\propto z^{2}\exp{-(z/z_{0})},$
where the normalization is fixed so that $\int_{0}^{\infty}n(z)\dd z = N_{\mathrm{A}},$ the mean surface density of 
sources in the survey. For weak lensing studies of dark energy, the most ambitious planned experiments for the next ten years will be LSST and Euclid; these surveys correspondingly have the most stringent calibration requirements, and so we find it useful to quantitatively phrase our results for future very-wide-area surveys such as these. Thus unless stated otherwise we 
choose $\na=30 \,\mathrm{gal/arcmin^2},$ and $z_{0}=0.34,$ corresponding to a median redshift of unity. We emphasize here, though, that the qualitative trends in all of our results are unchanged by the particular details of the survey characteristics.

We treat photometric redshift uncertainty in a relatively general manner, following the previous work of Ref.~\cite{ma_etal06}.  
We assume that the source galaxies are binned according to {\em photometric} redshift and that the 
true redshift distributions of the galaxies within each of the photometric redshift bins, $\niz$, 
are related to the overall, true galaxy source redshift distribution, $n(z)$, via 
$$\niz=n(z)\int_{z^{i}_{\mathrm{low}}}^{z^{i}_{\mathrm{high}}}\dd\zphot\phzpdf,$$ 
where ${z^{i}_{\mathrm{low}}}$ and ${z^{i}_{\mathrm{high}}}$ are the boundaries of the $i^{\mathrm{th}}$ photometric 
redshift bin.  Photo-z uncertainty is controlled by the function $\phzpdf.$  We take this to be a Gaussian at each redshift,
\beq
\label{eq:photozpdf}
\phzpdf = \frac{1}{\sqrt{2\pi}\sigma_{z}}\exp{\left[-\frac{(z - \zphot - \zbias)^{2}}{2\sigma_{z}^{2}}\right]}.
\eeq
This may seem to be overly restrictive, but as the mean, $\zbias$, and dispersion, $\sigma_z$, of this distribution 
can vary with redshift themselves, this parameterization allows for a wide variety of possible forms for the functions $\niz$.  
We adopt a fiducial model for photometric redshift error in which $\sigma_z = 0.05(1+z)$, while $\zbias=0$ at all redshifts.  
We use these fiducial functions to set the values of $\sigma_z$ and $\zbias$ at 31 control points, tabulated at intervals of $\Delta z=0.1$ 
between z=0 and z=3.  The values of $\zbias$ and $\sigma_z$ at each of these redshifts are free parameters in our forecasts, so that we 
model photo-z uncertainty with $2\times31=62$ free parameters.   This choice of binning allows for maximal degradation in dark energy 
constraint in the absence of prior information about the photometric redshift distribution of source galaxies (priors that would result from, for example, 
a photo-z calibration program).  This indicates that our parameterization 
does not enforce correlations that yield better-than-expected dark energy constraints.  
The dark energy constraints in the absence of prior information reduce to the same constraints 
that would be obtained with no binning in photometric redshift.

To model uncertainty in the fiducial photo-z distribution, we introduce priors on the values of the dispersion and bias at the $i^{\mathrm{th}}$ redshift control points, $\sigma^{i}_{z}$ and $\zbias^{i},$ respectively. These priors are 
\begin{eqnarray}
\label{eq:photozpriors}
\Delta\sigma^{i}_{z} = \sigma^{i}_{z}\sqrt{\frac{1}{2\nspeci}} \\
\label{eq:photozpriors2}
\Delta\zbias^{i} = \frac{\sigma^{i}_{z}}{\sqrt{\nspeci}},
\end{eqnarray}
where $\nspeci$ is the number of spectroscopic galaxies used in each of the 31 bins of width $\delta z = 0.1$ to calibrate the photo-z distribution. Our implementation of priors on the photo-z parameters is certainly simplistic. For example, we have further simplified our calculations by setting all of the $\nspeci$ equal to each other, effectively assuming that the calibrating spectra are sampled equally in redshift, whereas in practice there will be looser constraints on sources at high redshift than at low redshift. However, because the details of how a realistic calibration program will proceed remains uncertain at the present time, we use this simple model for prior information and postpone a refinement of this parameterization until the exact set of spectra that will be used to calibrate LSST and Euclid is better known. 

We emphasize here that $\nspec$ provides a convenient way to specify a one-parameter family of photo-z priors through Eqs.~\ref{eq:photozpriors} \& \ref{eq:photozpriors2}. The quantity $\nspec$ is likely not the true size of the spectroscopic calibration sample, but rather the equivalent size of a sample that fairly represents the color space distribution of the sources used in the lensing analysis.

%-------------------------------------------------------------------------------------------------------------------
\subsection{Observables}
\label{sub:obs}

We take the lensing power spectra of source galaxies binned by photometric redshift as 
well as the galaxy power spectra and galaxy-lensing cross spectra as observables that may 
be extracted from large-scale photometric surveys.  
The Limber approximation relates the power spectrum $\mathcal{P}_{s_{i}s_{j}}(k,z)$ associated 
with the correlation function of a pair of three-dimensional scalar fields, $\ssubi$ and $\ssubj,$ to its 
two-dimensional projected power spectrum, ${P}_{\xsubi\xsubj}(\ell):$
\beq
\label{eq:limber}
P_{\xsubi\xsubj}(\ell)=\int dz \frac{\wiz\wjz}{\dasqz H(z)}\mathcal{P}_{\ssubi\ssubj}(k=\ell/\da(z),z).
\eeq
Eq.~\ref{eq:limber} essentially describes how the two-dimensional scalar fields $\xsubi$ are observed as projections of
the three-dimensional scalar fields, $\ssubi.$ 
The angular diameter distance function is denoted by $\da,$ and $H(z)$ is the Hubble parameter.

The weight function $\wiz$ specifies the projection of the 3D source fields onto the 2D projected fields:
\beq
\label{eq:kernel}
\xsubi(\mathbf{\hat{\mathbf{n}}})=\int dz\wiz \ssubi(\da\mathbf{\hat{\mathbf{n}}},z).
\eeq
For galaxy fluctuations, the weight function is simply the redshift distribution
of galaxies in the $i^{\mathrm{th}}$ tomographic bin, $\niz,$ times the Hubble rate:
$$\mathrm{W}^{g}_{\mathrm{i}}(z)=H(z)\niz.$$
The weight function associated with fluctuations in lensing convergence is given by
$$W^{\kappa}_{\mathrm{i}}(z)=\frac{3}{2}H_{0}^{2}(1+z)\Omega_{\mathrm{m}}\daz\int_{z}^{\infty} dz'\frac{\da(z,z')}{\da(z')}n_{\mathrm{i}}(z'),$$
where $\da(z,z')$ is the angular diameter distance between $z$ and $z'.$
 
In principle, neither the redshift distribution of the galaxies used for the galaxy clustering nor 
the tomographic binning scheme need be the same as that used for cosmic shear sources, but for simplicity 
we use the same underlying distribution and binning for both so that the chief difference between the 
galaxy power spectrum $\pgigj,$ the convergence power spectrum $\pkikj,$ and the cross-spectrum $\pkigj,$ 
is the form of the weight functions.   Above and throughout, lower-case Latin indices label the tomographic redshift 
bin of the sources.  For a survey with its galaxies divided into $N_g$ redshift bins used to measure the galaxy 
clustering, and $N_s$ bins for the galaxies used to measure cosmic shear, there will be $N_g(N_g+1)/2$ distinct 2-D 
galaxy power spectra $\pgigj,$ $N_s(N_s+1)/2$ distinct convergence power spectra $\pkikj,$ 
and $N_{s}N_{g}$ distinct cross-spectra $\pkigj.$

The matter power spectrum, $\mathcal{P}_{\ssubi\ssubj}(k,z)=\mathcal{P}_{\delta}(k,z),$ sources the 
three-dimensional power in cosmic shear, whereas the source power for galaxy-galaxy correlations is the 3-D galaxy power spectrum
$\mathcal{P}_{\ssubi\ssubj}(k,z)=\mathcal{P}_{\gi\gj}(k,z).$
In all of our calculations we restrict galaxy correlation information to low multipoles $\ell\leq300;$ at redshift z=1 this corresponds 
to fluctuations of wavenumber $k\approx0.2\hmpcinv,$ so it will suffice
for our purposes to use a linear, deterministic bias to relate the mass overdensity, $\delta(z),$ to the galaxy overdensity, $\delta_{g}(z)=b(z)\delta(z).$  

We allow for a very general redshift-dependent bias. To model uncertainty in the galaxy bias function $b(z),$ we allow the bias to vary freely about its fiducial value of unity in $N_b$ galaxy bias bins, evenly spaced in true redshift, so that uncertainty in galaxy bias is encoded by $N_b$ parameters. We computed dark energy constraints using the Fisher analysis technique described in~\S\ref{sub:forecasting} for $N_b$ ranging from $1$ to $30.$ We find that the dark energy constraints are insensitive to $N_b$ ranging from $1-15.$ Throughout this manuscript, we present results pertaining to $N_b=10$ bins, so that the value of the galaxy bias function $b(z)$ has independent, parametric freedom in redshift bins of width $\delta z=0.3.$ While finer binning is possible, particularly if the number of tomographic galaxy bins $N_g$ is increased, a further increase of parametric freedom is unnecessary as galaxy bias is not a rapidly varying function of redshift (see, for example, Ref.~\cite{coil_etal04}). 

As a further simplification, we set the {\em fiducial value} of the bias function to unity at all redshifts, $\biz=b(z)=1$ at all $z$ for all $i.$  This choice of fiducial parameter values is conservative, 
because the galaxies observed as part of high redshift samples will likely be biased \cite{coil_etal06}, exhibiting relatively 
stronger correlations than matter, so we underestimate signal-to-noise of galaxy clustering measurements 
in the fiducial case. 
%We discuss how we model uncertainty in galaxy bias in~\S\ref{sub:forecasting}.

%-------------------------------------------------------------------------------------------

\subsection{Parameter Forecasting}
\label{sub:forecasting}

We estimate the constraints from upcoming photometric surveys using the formalism of the Fisher information matrix.  
Useful references for this formalism include \cite{jungman_etal96,tegmark_etal97,seljak97,kosowsky_etal02,albrecht_etal06}. 
The Fisher matrix is defined as 

\beq
\label{eq:fisher}
F_{\alpha \beta}=\sum_{\ell_{\mathrm{min}}}^{\ell_{\mathrm{max}}} (2\ell+1)\fsky 
\sum_{\mathrm{A,B}} \frac{\partial \mathcal{O}_{\mathrm{A}}}{\partial p_{\alpha}} 
\mathbf{C}^{-1}_{\mathrm{AB}} 
\frac{\partial \mathcal{O}_{\mathrm{B}}}
{\partial p_{\beta}} + F_{\alpha \beta}^{\mathrm{P}}. 
\eeq
The parameters of the model are $p_{\alpha}$ and the 
$\mathcal{O}_{\mathrm{A}}$ are the observables described in \S~\ref{sub:obs}.  
Greek indices label model parameters while Latin, upper-case indices label 
distinct observables.  We take $\ell_{\mathrm{min}}=2$ for all observables.  For the 
lensing spectra, we take $P_{\kappa\kappa}$ observables we set $\ell_{\mathrm{max}}=3000$ so that 
the assumptions of weak lensing and Gaussian statistics remain relatively reliable 
\cite{white_hu00,cooray_hu01,vale_white03,dodelson_etal06,semboloni_etal06}.  
For the galaxy clustering statistics, we eliminate small-scale information so that we do not 
need to model scale-dependent galaxy bias, which is potentially complicated in itself.  
Therefore, we set $\ell_{\mathrm{max}}=300$ for $P_{gg}$ and $P_{\kappa g}$, corresponding roughly to angular scales of $\sim1$ degree. 
We emphasize that this restriction is very conservative as it implies that our joint analysis does {\em not} employ the use of Baryon Acoustic Oscillation features in our galaxy power spectra. As we will see, the added benefit of a joint analysis stems primarily from the increased ability to self-calibrate parameterized uncertainty in $\pmatk$ and the distribution of photometric redshifts.

In Eq.~\ref{eq:fisher}, $\mathbf{C}^{-1}_{\mathrm{AB}}$ is the inverse of the covariance matrix; our treatment of the  
covariance matrix calculation, and its associated Fisher matrix, is very similar to that in Ref.~\cite{hu_jain04}, to which we refer the reader for additional details.  
Briefly, the covariance between a pair of power spectra $P_{\xsubi\xsubj}$ and $P_{\xsubm\xsubn}$
is given by
\beq
\label{eq:covone}
\mathrm{\mathbf{Cov}}(P_{\xsubi\xsubj},P_{\xsubm\xsubn})
=\tilde{P}_{\xsubi\xsubm}\tilde{P}_{\xsubj\xsubn} + \tilde{P}_{\xsubi\xsubn}\tilde{P}_{\xsubj\xsubm},
\eeq
where in the case of either galaxy power or convergence power the observed spectra $\tilde{P}_{\xsubi\xsubj}$ 
have a contribution from both signal and shot noise,
 $$\tilde{P}_{\xsubi\xsubj}(\ell)=P_{\xsubi\xsubj}(\ell) + \mathrm{N}_{\xsubi\xsubj},$$
where $\mathrm{N}_{\gi\gj}=\delta_{\mathrm{ij}}\mathrm{N}_{\mathrm{i}}^{\mathrm{A}}$ 
is the shot noise term for galaxy spectra,
with $\mathrm{N}_{\mathrm{i}}^{\mathrm{A}}$ denoting the surface density of sources, 
and $\mathrm{N}_{\kappai\kappaj}=\delta_{\mathrm{ij}}\gamma_{\mathrm{int}}^{2}\mathrm{N}_{\mathrm{i}}^{\mathrm{A}}$ 
is the shot noise for convergence.  We calculate the observed cross-spectra $\tilde{P}_{\kappai\gj}$ without a contribution from 
shot noise, so that $\tilde{P}_{\kappai\gj}=P_{\kappai\gj}$, because galaxies are lensed by mass separated from them by 
cosmological distances, so the cross-correlation of the noise terms should be small.  
We adopt a common convention of setting the intrinsic galaxy shape noise $\gamma_{\mathrm{int}}=0.2$
and absorb differences in shape noise between different surveys into the surface density of sources $\na.$

The inverse of the Fisher matrix is an estimate of the parameter covariance near 
the maximum of the likelihood, i.e. at the fiducial values of the parameters.  
Therefore, the measurement error on parameter $\alpha$ marginalized 
over all other parameters is
%
%-- error
\beq
\label{eq:ferror}
\sigma(p_{\alpha})=\sqrt{[F^{-1}]_{\mathrm{\alpha \alpha}}}.
\eeq
Gaussian priors on the parameters are incorporated into the Fisher analysis via $F_{\alpha \beta}^{\mathrm{P}}$ 
in Eq.~\ref{eq:fisher}.  If one is instead interested in unmarginalized errors, for example to test the intrinsic sensitivity of 
the observables to a parameter $p_{\alpha},$ the quantity $\sqrt{1/[F]_{\mathrm{\alpha \alpha}}}$ provides an estimate of 
the uncertainty on $p_{\alpha}$ in the limit of zero covariance between $p_{\alpha}$ and any of the other parameters 
in the analysis.

The Fisher formalism can also be used to estimate the magnitude of a bias that would occur in parameter inference 
due to a systematic error in the observables.  If $\Delta\mathcal{O}_{\mathrm{A}}$ denotes the difference between the fiducial 
observables and the observables perturbed by the presence of the systematic error, then the systematic offset in the inferred 
value of the parameters caused by the error can be estimated as 
\beq
\label{eq:fishersystematic}
\delta p_{\alpha} = \sum_{\beta} [F^{-1}]_{\alpha \beta} 
\sum_{\ell} (2\ell+1) \fsky \sum_{\mathrm{A,B}} 
\Delta \mathcal{O}_{\mathrm{A}}\mathbf{C}^{-1}_{\mathrm{AB}} 
\frac{\partial \mathcal{O}_{\mathrm{B}}}{\partial p_{\beta}}.
\eeq

We assume a standard, flat $\Lambda$CDM cosmological model and vary 
seven cosmological parameters with fiducial values are as follows: 
$\Omegam h^2= 0.13,$  $\wzero=-1,$ $\wa=0,$ $\Omegab h^2 = 0.0223,$ $\ns=0.96,$ 
$\mathrm{ln}(\cmbnorm)=-19.953,$ and $\Omega_{\Lambda}=0.73.$  We utilize the following marginalized 
priors: $\Delta\Omegam h^2 = 0.007,$ $\Delta\Omegab h^2 = 0.001,$ $\Delta\ns=0.04 ,$ $\Delta\mathrm{ln}(\cmbnorm) = 0.1.$  
These priors are comparable to contemporary uncertainty \cite{komatsu_etal11}, so this choice should be conservative. We have verified that strengthening these priors to levels of uncertainty that will be provided by Planck \cite{hu_etal06} does not induce a significant change to any of our results.

We determined the number of independent parameters for the matter power spectrum with 
an analysis of the off-diagonal elements of the inverse Fisher matrix.  The parameter covariance 
is
\beq
\label{eq:paramcovdef}
\qab\equiv\frac{\fishinvab}{\sqrt{\fishinvaa\fishinvbb}}.
\eeq
In general, $-1\leq\qab\leq1,$ with $\qab = (-)1$ corresponding to the case where parameters $\para$ and $\parb$ are perfectly (anti-)correlated and 
$\qab=0$ corresponding to uncorrelated parameters. By increasing the number of matter power spectrum parameters until the Fisher matrix is no longer invertible in the absence of prior information on these parameters, we determined that a lensing-only analysis reaches a level of total information loss when ten $\dpmati$ parameters are used. By studying the behavior of the off-diagonal Fisher Matrix entries as the number of matter power spectrum parameters are increased, we find that this state of information loss occurs after values of $\qab=\pm0.8$ obtain between pairs of distinct $\dpmati,$ in agreement with the method used in Ref.~\cite{huterer_takada05} to arrive at this conclusion (D.~Huterer, private communication). Because galaxy correlation observables provide additional information with which to self-calibrate matter power spectrum parameters, including $\pgg$ and $\pkg$ allows for slightly finer binning in wavenumber, 
but for the sake of facilitating a direct comparison between the different sets of observables we have limited our analysis to ten parameters $\dpmati,$ irrespective of whether we consider a joint analysis or weak lensing alone.

There is additional freedom in the choice of the number of tomographic bins one uses to divide both the sources used to measure lensing as well as the sources used to measure galaxy correlations. As has been noted in previous studies, for example Ref.~\cite{ma_etal06}, dark energy information from lensing saturates at $N_s=5$ tomographic bins; this saturation point is determined by using the Fisher matrix (in the limit of perfect prior knowledge of all nuisance parameters, in our case the photo-z parameters $\sigma^{i}_{z}$ and $z^{i}_{\mathrm{bias}},$ and the $\dpmati$ parameters) to compute the statistical constraints $\sigma(\wzero,\wa)$ and increasing the number of tomographic bins until the constraints cease to improve. We find that for the case of galaxy clustering this information saturation occurs at $N_g=10$ tomographic bins, although we note that this saturation point depends on the maximum multipole used in the analysis. For example, 
Ref.~\cite{zhan_etal08} uses $\ell_{\mathrm{max}}=2000$ for their galaxy clustering analysis and finds that additional information is available by increasing the number of bins to $N_g=40.$ Our smaller $N_g$ saturation point is a consequence of our conservative choice for $\ell_{\mathrm{max}},$ and the lack of BAO information implied by this choice.

%----------------------------------------------------------------------------------------------------------------------------------------------
\section{Results}
\label{section:results}

%----------------------------------------------------------------------------------------------------------------------------------------------
\subsection{Power Spectrum Self-Calibration}
\label{subsection:pkscale}

We begin by presenting our calculation of the scale-dependence of the sensitivity 
of weak lensing (with and without galaxy correlations) to the matter power spectrum.  
Our second model for $\pmatk$ uncertainty is well-suited to this investigation: the constraints 
on parameter $\dpmati$ provide an estimate of the statistical significance of the weak lensing 
signal produced by correlations in the matter distribution on scales $k\approx k_{\mathrm{i}}.$  
In the prevailing jargon, this calculation corresponds to {\em self calibration} of the matter power 
spectrum.  

\begin{figure*}[t!]
\centering
\includegraphics[width=12.0cm]{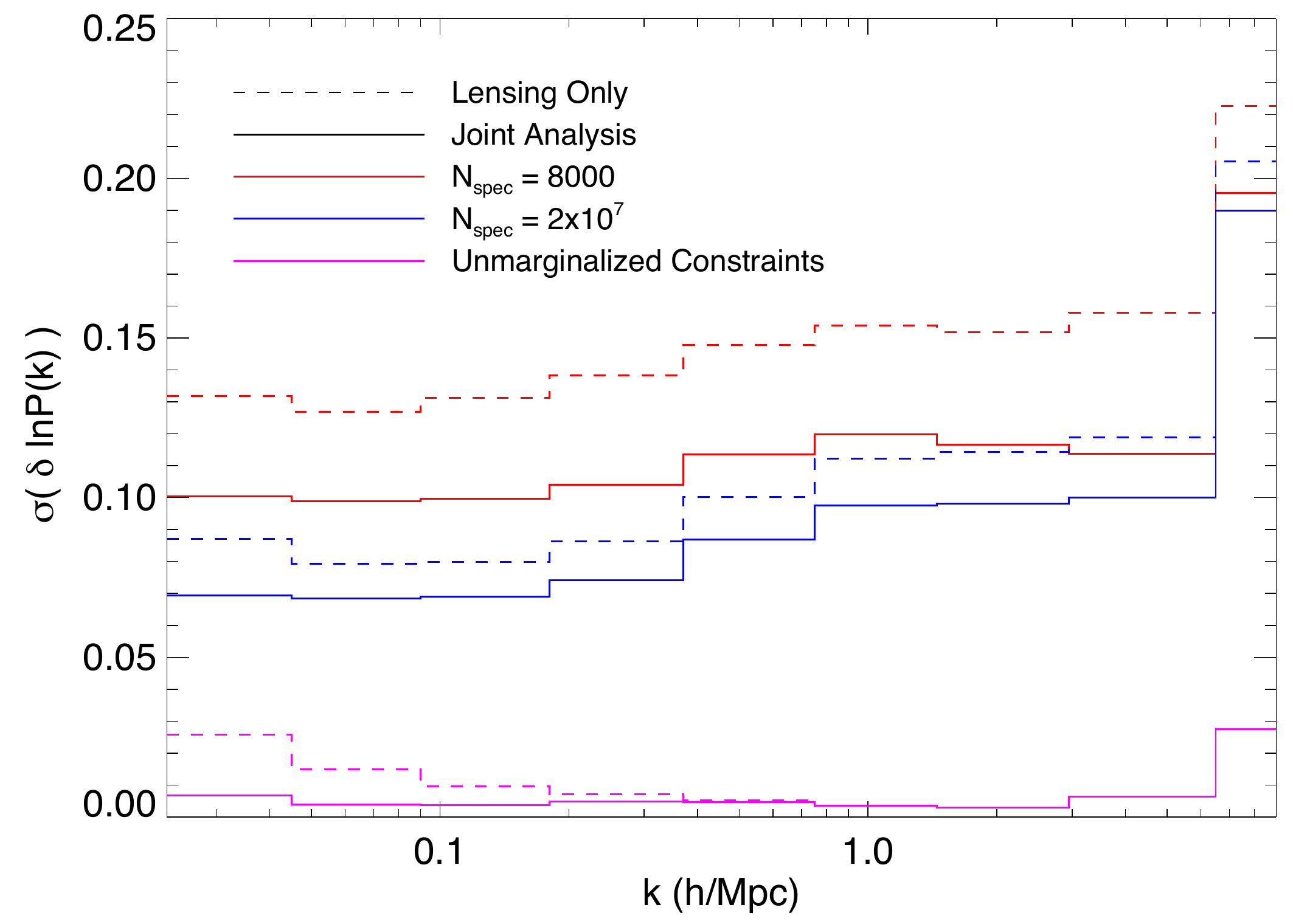}
\caption{
Statistical constraints on $\dpmati,$ the parameters encoding uncertainty in the calibration of the matter power spectrum, are plotted against the scale of the wavenumber. We plot unmarginalized constraints as magenta curves; the minimum of the magenta curves at $k\approx2$ $\hmpcinv$ illustrates that the intrinsic sensitivity of the weak lensing signal peaks at this scale. With the red and blue curves we plot marginalized constraints on the $\dpmati$ parameters for different levels of uncertainty in the distribution of photometric redshifts. The vertical axis values for the red and blue curves give the statistical precision with which a future very-wide-area survey such as LSST or Euclid will be able to self-calibrate the theoretical prediction for the matter power spectrum on a given scale.
}
\label{fig:ht1a}
\end{figure*}

The constraints from this computation are plotted as a function of scale in Fig.~\ref{fig:ht1a}.
The magenta curves at the bottom of the Fig.~\ref{fig:ht1a} pertain to {\em unmarginalized} constraints on the $\dpmati.$  
In other words, covariant uncertainty in cosmology, photometric redshifts, and galaxy bias is {\em not} 
taken into account in the magenta curves.  In plotting the red and blue curves we illustrate our results when this covariance 
is accounted for by marginalizing over all other parameters in the analysis.  The red curves correspond to a calculation with $\nspec=8000,$ or $\Delta\sigma_z/\sigma_z\approx10^{-2}.$ The blue curves pertain to a calculation with $\nspec=2\times10^{7},$ which is sufficiently large that further increases to $\nspec$ do not improve constraints on any of the parameters in our analysis, so priors this tight effectively correspond to the case where the photo-z distribution is known perfectly. The solid curves include galaxy correlation observables ($\pgg$ and $\pkg$) in 
addition to lensing observables and thus lie strictly below the dashed curves (lensing only). 
The step-like appearance of the curves reflects the coarse binning in wavenumber of our parameterization of $\pmatk$ uncertainty: 
a forthcoming very-wide-area, LSST- or Euclid-like survey is only able to constrain $\sim 10$ independent matter power spectrum 
parameters (see \S~\ref{sub:forecasting}).  While each of the curves in Fig.~\ref{fig:ht1a} pertains to a calculation in which we used 
the Smith et al.~\cite{smith_etal03} fitting formula as our fiducial $\pmatk,$ the results using either the halo model or the Peacock \& Dodds \cite{peacock_dodds96} 
fitting formula are nearly identical, so conclusions drawn from Fig.~\ref{fig:ht1a} are quite robust to detailed changes in the fiducial model 
for nonlinear collapse.

The minimum of the unmarginalized constraints in Fig.~\ref{fig:ht1a} at $k \sim2 \hmpcinv$ occurs on the scale at which 
weak lensing is most intrinsically sensitive to matter overdensities.
This minimum occurs on a physical scale nearly an order of magnitude smaller than the minimum of the 
marginalized constraints. This observation is an extension of the previous work of Ref.~\cite{huterer_takada05} 
and is itself an important result as it demonstrates the need for precision in the prediction for $\pmatk$ over the full range of nonlinear scales $k\lesssim5$ $\hmpcinv.$  Because this shift in scale-dependence occurs even for the case of perfect prior knowledge on the photo-z parameters $(\nspec=2\times10^7),$ then it is not the effect of photometric redshift uncertainty 
that drives this shift in scale, but rather degeneracy with cosmological parameters. An analysis of the off-diagonal Fisher matrix elements shows that covariance of the $\dpmati$ with dark energy parameters is chiefly responsible for this dramatic shift in the scale-dependence. To be specific, 
 with $\pari=\dpmati$ and $\parj=\wzero,\wa,$ the corresponding $\qij$ (Eq.~\ref{eq:paramcovdef}) have the maximum magnitudes of any of the cosmological parameters in our parameter set and they attain 
their maxima at $k\approx0.1\hmpcinv.$
%---------------------------------------------------------------------------------------------------
%-----------------------------------------------------------------------------------------------------

%------------------------------------------------------------------------------
\subsection{Statistical constraints on Dark Energy}
\label{subsection:decons}

We proceed with results on the sensitivity of dark energy constraints to uncertainty in predictions of the 
nonlinear evolution of $\pmatk$, incorporating possible additional uncertainty from photometric 
redshift errors.  Results for the $\dpmati$ 
model appear in \ref{subsubsection:decons_dpmati} while those pertaining to the 
more restrictive `halo model" treatment of power spectrum uncertainty are 
discussed in \ref{subsubsection:decons_hm}.

%-------------------------------------------------------------------
\subsubsection{The $\dpmati$ Model}
\label{subsubsection:decons_dpmati}

%---------------------------------------------------------------------------------------------------
\begin{figure*}[!th]
\centering
\includegraphics[width=10.4cm]{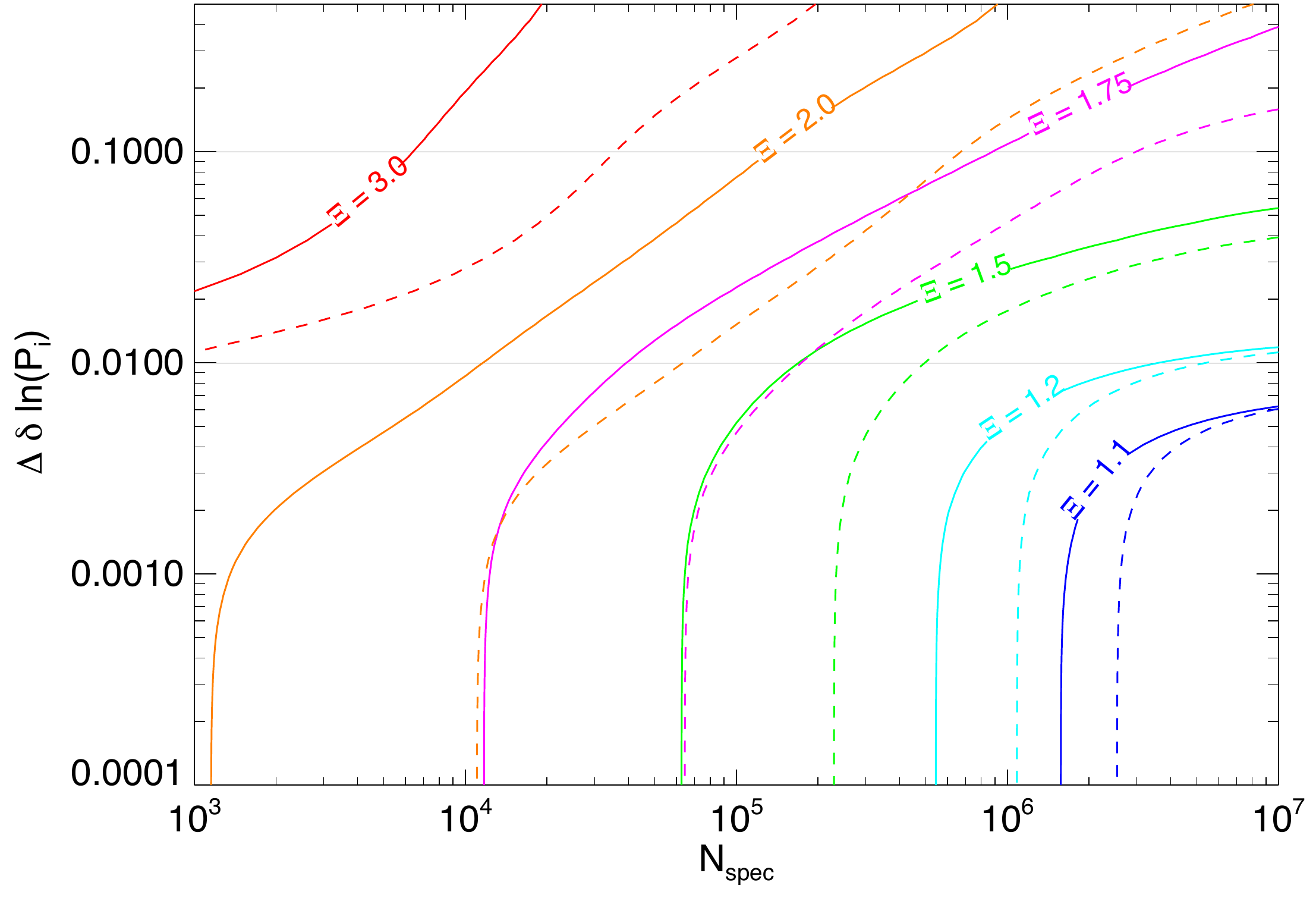}
\includegraphics[width=10.4cm]{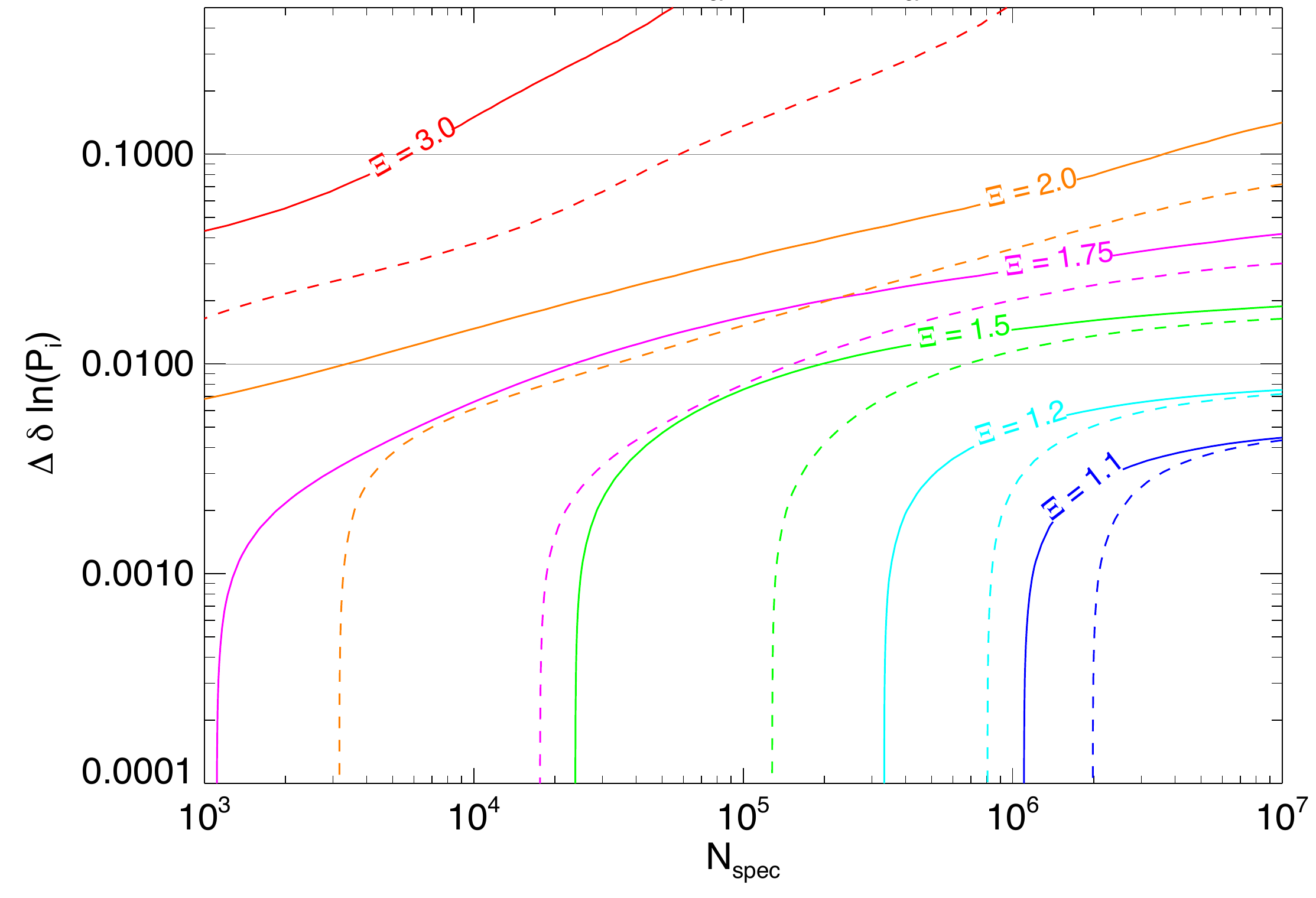}
\caption{
Contours of degradation in $\wzero$ and $\wa$ constraints appear in the top and bottom panels, respectively. 
The degradation is quantified by $\Xi\equiv\sigma(\wzero, \wa)/\sigma^{\mathrm{perf}}(\wzero, \wa),$ 
where $\sigma^{\mathrm{perf}}$ is the level of statistical uncertainty of an LSST- or Euclid-like survey in the limit of 
perfect certainty on the photo-z distribution and the nonlinear evolution of $\pmatk.$ 
The precision in the calibration of the matter power spectrum $\pmatk$ appears on the vertical axes, while the priors on 
the photo-z distribution, parameterized by $\nspec,$ appear on the horizontal axes.  Prior information about the 
functions $\zbias$ and $\sigma_z$, which govern the uncertainty in the photo-z distribution $\phzpdf$, is distributed uniformly in 
redshift according to $\Delta\zbias=\sqrt{2}\Delta\sigma_z=\sigma_z/\sqrt{\nspec} $.  Priors on the $\dpmati$ parameters are 
distributed uniformly in among bandpowers in $\mathrm{log}(k).$
Dashed curves pertain to a survey using only weak lensing information, solid curves to a joint analysis that includes galaxy clustering. 
The gray, horizontal lines roughly bound the range of matter power spectrum uncertainty that is attainable in advance of these surveys.  
}
\label{fig:wcons_dpmati}
\end{figure*}
%-----------------------------------------------------------------------------------------------------

In Figure~\ref{fig:wcons_dpmati} we depict contours of the degradation in the statistical constraints on the dark energy equation of state 
associated with simultaneous uncertainty in $\pmatk$ and photometric redshifts.  The constraints on $\wzero$ and $\wa$ are shown in 
units of the perfectly calibrated limit, when power spectra and photometric reshifts are known so well as to be inconsequential to the dark 
energy error budget; we denote the constraints on $\wzero$ and $\wa$ in the limit of perfect calibration by $\sigma^{\mathrm{perf}}(\wzero)$ and $\sigma^{\mathrm{perf}}(\wa),$ respectively.  For example, we plot the ratio 
$\Xi=\sigma(\wzero)/\sigma^{\mathrm{perf}}(\wzero)$ to illustrate the level of degradation of $\wzero$ constraints. 
For the sake of scaling our results to an absolute statistical constraint, we summarize these baseline constraints in Table \ref{table:constraints}.
We refer the reader to \S~\ref{section:conclusions} for a discussion of how the results we present here change when considering a survey with characteristics similar to DES.

%%%%%%%%%%%%%%%%%%%%%%% TABLE 1  %%%%%%%%%%%%%%%%%%%%%%%%%%%%%%%%%%%

\begin{table}
\caption{Baseline Constraints}
\vspace*{-12pt}
\begin{center}
\begin{tabular}{lcccccccr}
\hline\hline
\vspace*{-8pt}
\\
\multicolumn{1}{c}{Observables}&
\multicolumn{1}{c}{$\sigma(\wzero)\ $} &
\multicolumn{1}{c}{$\sigma(\wa)\ $} &
\multicolumn{1}{c}{$\sigma(w_{\mathrm{piv}})\ $}
\\
\hline\hline
\\
Weak Lensing Only &  $ 0.071 $ & $ 0.22 $ & $ 0.022$  \\
Joint Analysis & $ 0.058 $ & $ 0.19 $  & $ 0.018 $ \\
\hline
\end{tabular}
\end{center}
{\sc Notes.}--- 
Column (1) specifies whether or not observables employing galaxy clustering ($\pgg$ and $\pkg$) were used in the calculation.   
Columns (2), (3), and (4) give the statistical constraints on $\wzero,$ $\wa,$ and $\wpiv,$ respectively, that can be obtained by future very-wide-area
surveys such as LSST or Euclid. Note that these constraints account for statistical errors only and so represent the optimistic 
limit of achievable dark energy constraints for a survey with these characteristics.

\label{table:constraints}
\end{table}

%%%%%%%%%%%%%%%%%%%%%%%%%%%%%%%%%%%%%%%%%%%%%%%%%%%%%%%%%%%%%%%%%%

The levels of constraint degradation depend upon the precision of both the matter power spectrum, parameterized by $\dpmati$, and 
the photometric redshift distributions of sources.  We quantify the precision of power spectrum prediction by a prior constraint 
on the $\dpmati$ parameters, $\Delta \dpmati$.  For example, a value of $\Delta \dpmati = 0.1$ corresponds to a 10\% precision 
on the bandpower at wavenumber $k_{\mathrm{i}}$.  For simplicity, we apply the same prior at all wave bands in order to 
produce Fig.~\ref{fig:wcons_dpmati}.  The assumed level of calibration of photometric redshifts is specified by the 
parameter $\nspec,$ as discussed in \S~\ref{sub:photoz}.

The horizontal gray lines roughly bound the range of precision in 
the prediction for $\pmatk$ on scales relevant to lensing and large-scale galaxy 
clustering that may be attainable by near-future numerical 
simulation campaigns. The precise values of the realized precisions will depend upon the resources dedicated to address 
this issue as well as the ability of data to constrain baryonic processes that alter 
power spectra and numerical simulations to treat these baryonic processes. 
For example, the Coyote Universe simulation campaign \cite{heitmann_etal09} has already achieved 
a $1\%$ calibration of the matter power spectrum to scales as small as $k\approx1$ $\hmpcinv;$ future results from, e.g.,
 the Roadrunner Universe \cite{habib_etal09} will improve upon these results, although it is still not clear in detail 
 how precisely $\pmatk$ will be calibrated to scales as small as $k\approx10$ $\hmpcinv,$ especially when uncertainty in 
 baryonic physics is taken into account.

Fig.~\ref{fig:wcons_dpmati} contains contours for parameter degradation when using lensing data alone (dashed curves) as 
well as the corresponding constraint degradations when both lensing and galaxy clustering observables are considered (solid curves). 
To be sure, the additional information available to a joint analysis guarantees that 
$\sigma^{\mathrm{perf}}(\wzero,\wa)$ from the joint galaxy clustering and lensing analysis is 
less than the corresponding constraint when considering lensing observables alone.  In each case,  
we plot the degradation $\Xi = \sigma(\wzero,\wa)/\sigma^{\mathrm{perf}}(\wzero,\wa)$ relative to the 
idealized constraints for that technique. 
Accounting for this difference, Fig.~\ref{fig:wcons_dpmati} highlights the dramatic relaxation on the calibration 
requirements of both power spectrum predictions and photometric redshifts provided by including galaxy clustering 
statistics.  For the purpose of studying dark energy, this relaxation is the most significant advantage provided by 
utilizing galaxy correlations in the analysis.  

To further illustrate this point, consider particular examples that can be gleaned from Fig.~\ref{fig:wcons_dpmati}.  
In the limit of perfect prior knowledge of $\pmatk$ and the photo-z distribution, the 
increase in constraining power that a joint analysis has over an analysis that includes lensing 
observables only is merely $\sim15-20\%$; however, as the precision in the calibration of both photometric redshifts 
and the matter power spectrum decreases, the dark energy constraints degrade by a factor of $2-5$ more rapidly 
when galaxy clustering information is neglected.  For example, suppose that predictions for the matter power spectrum 
attain $1\%$ level of precision in advance of LSST or Euclid. A joint analysis with the statistical equivalent of $\nspec \approx 20,000$ 
(a realistic proposition as we are assuming these calibrating galaxies to be distributed evenly in redshift) 
yields constraints on $\wa$ that are weakened by $75\%$ due to photometric redshift uncertainty, whereas if galaxy clustering 
information is neglected $\nspec \approx 100,000$ will be required to protect against the same level of degradation in dark 
energy constraints. 

\subsubsection{The Halo Model}
\label{subsubsection:decons_hm}

%---------------------------------------------------------------------------------------------------
\begin{figure*}[th!]
\centering
\includegraphics[width=7.0cm]{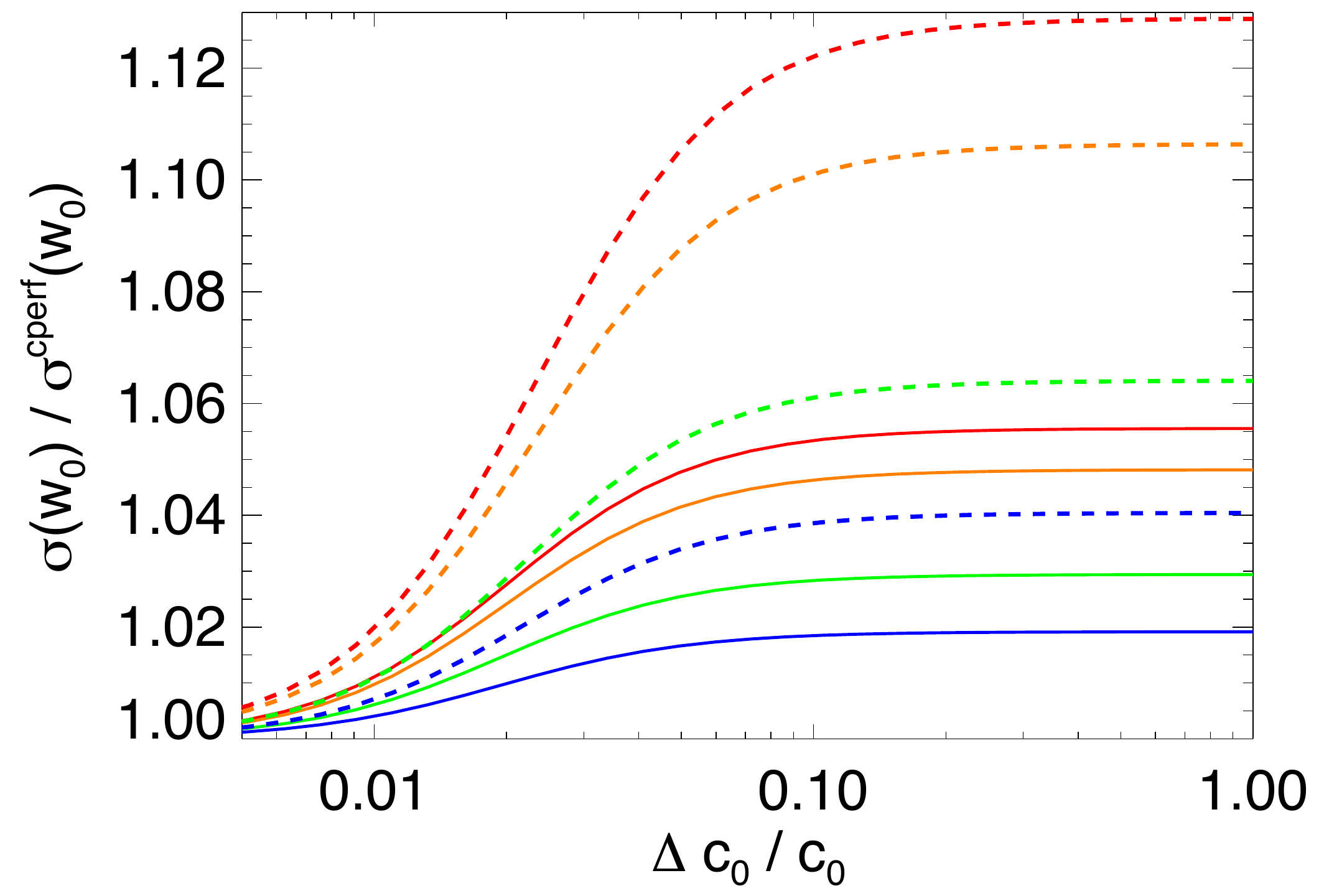}
\includegraphics[width=7.0cm]{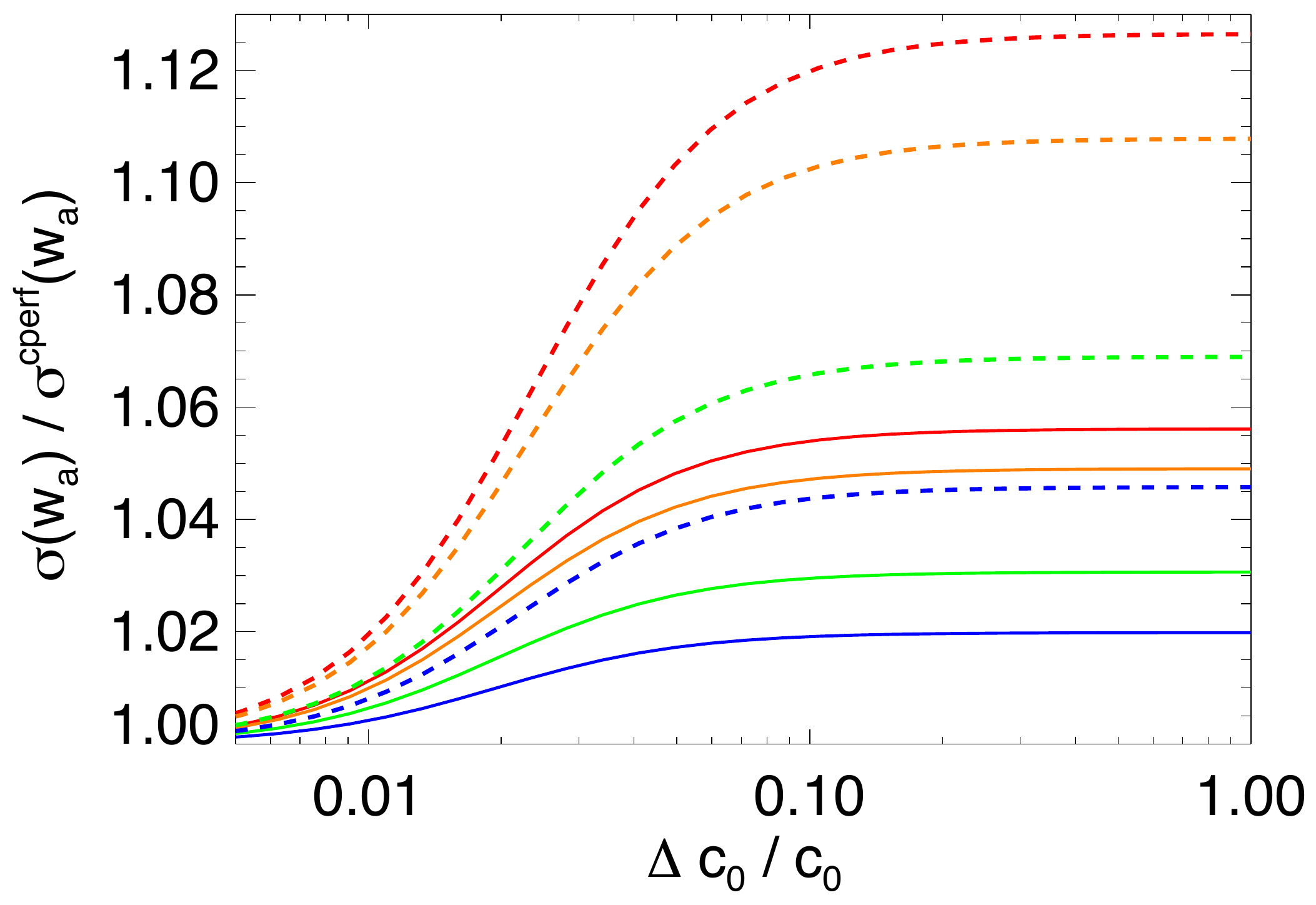}
\includegraphics[width=7.0cm]{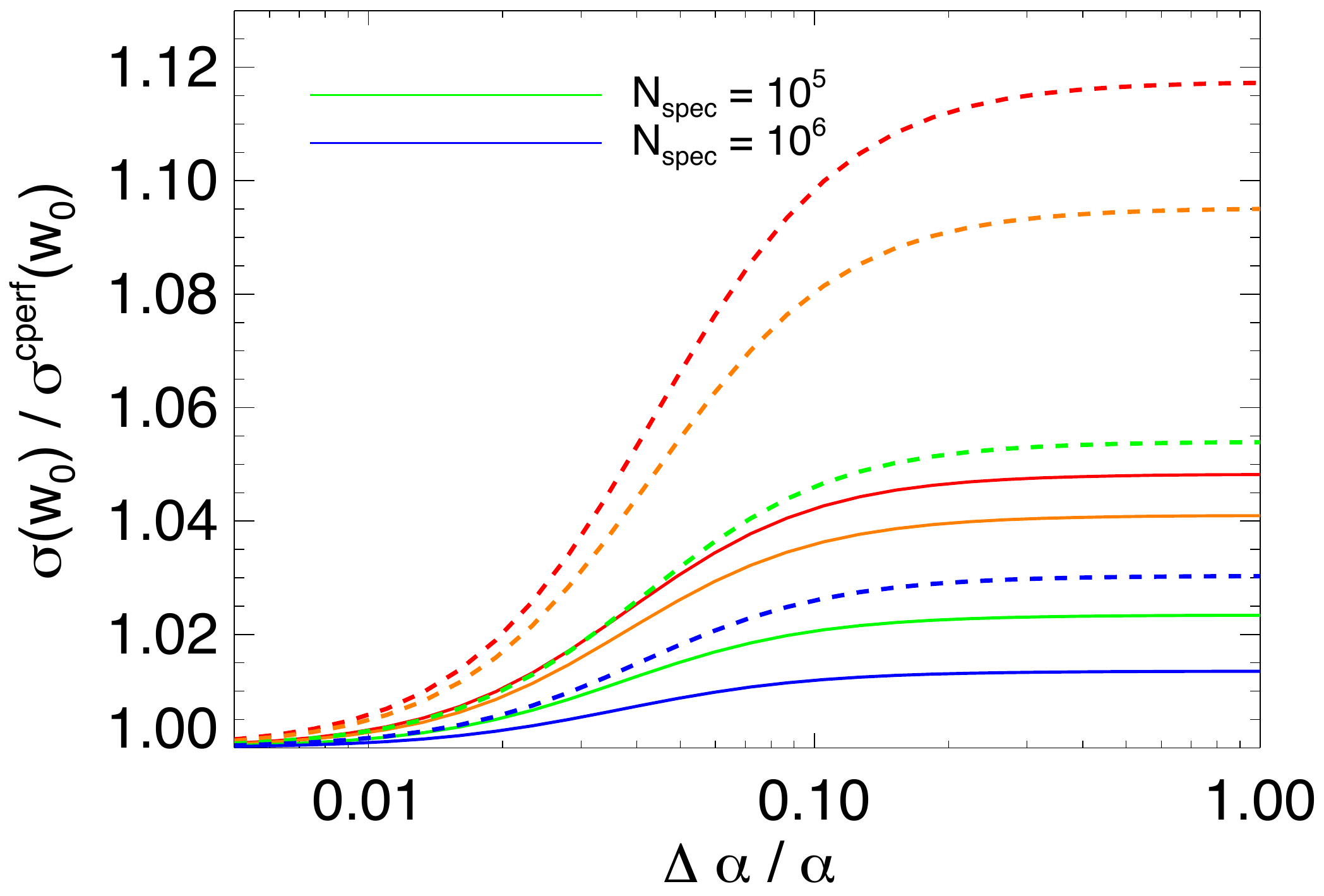}
\includegraphics[width=7.0cm]{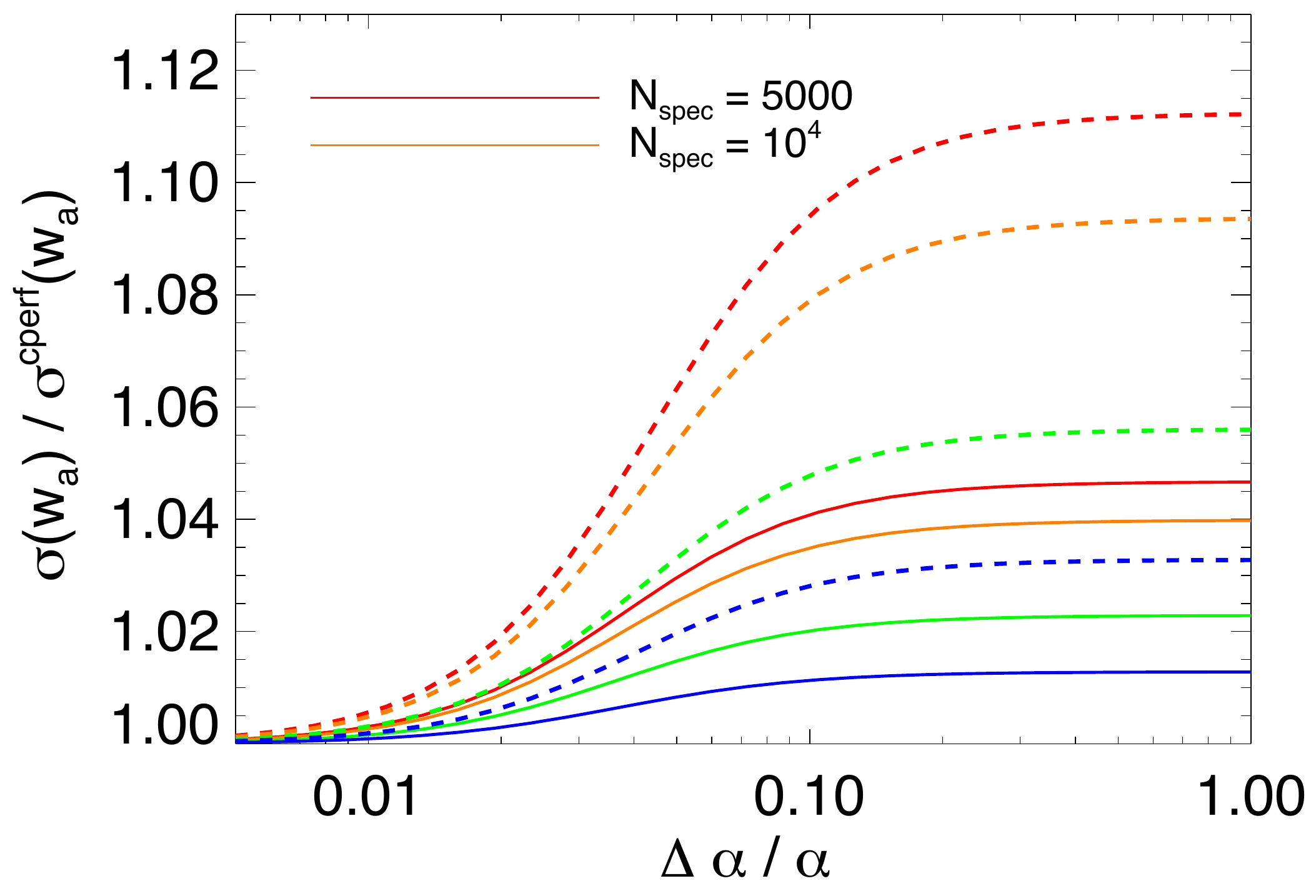}
\includegraphics[width=7.0cm]{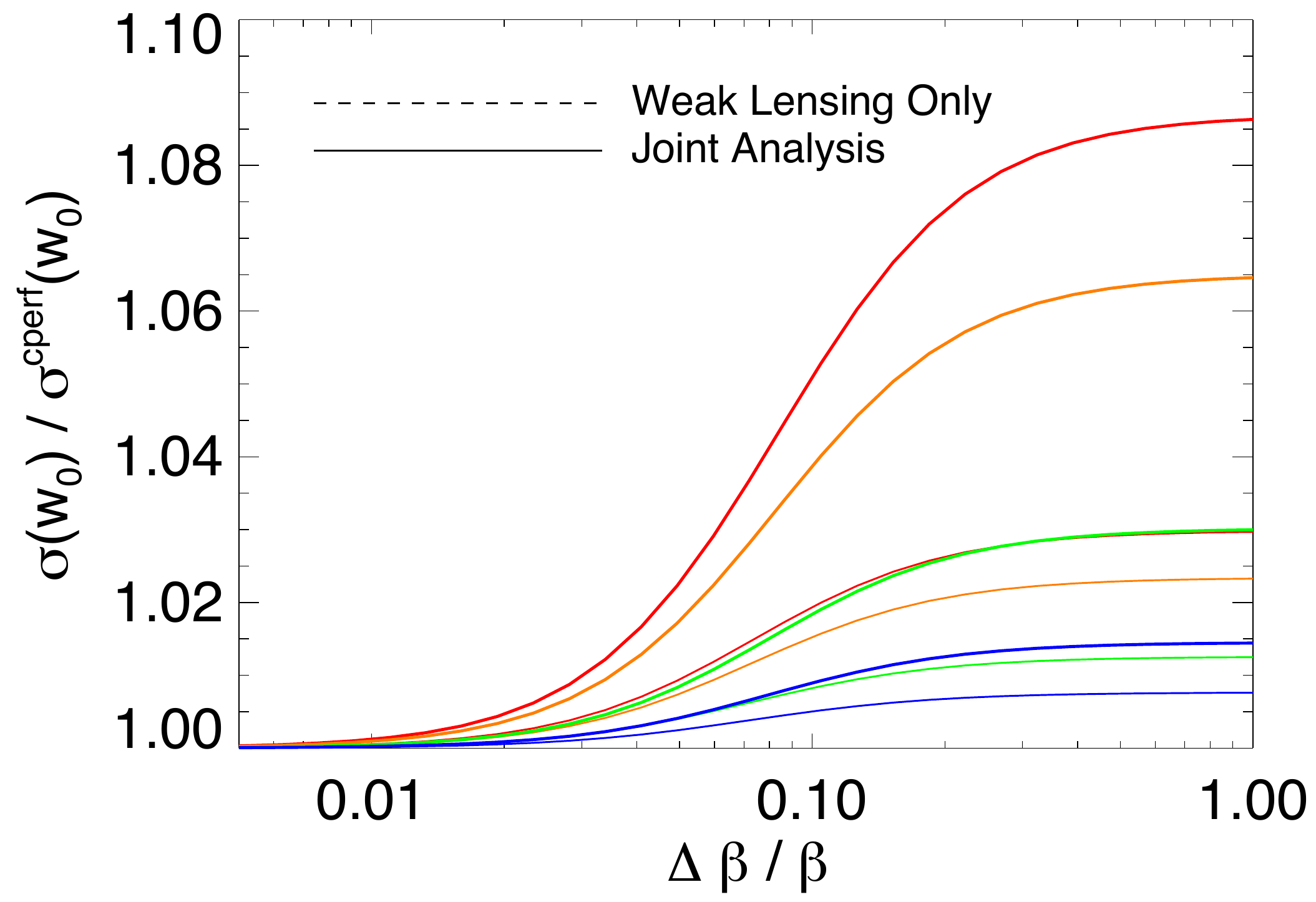}
\includegraphics[width=7.0cm]{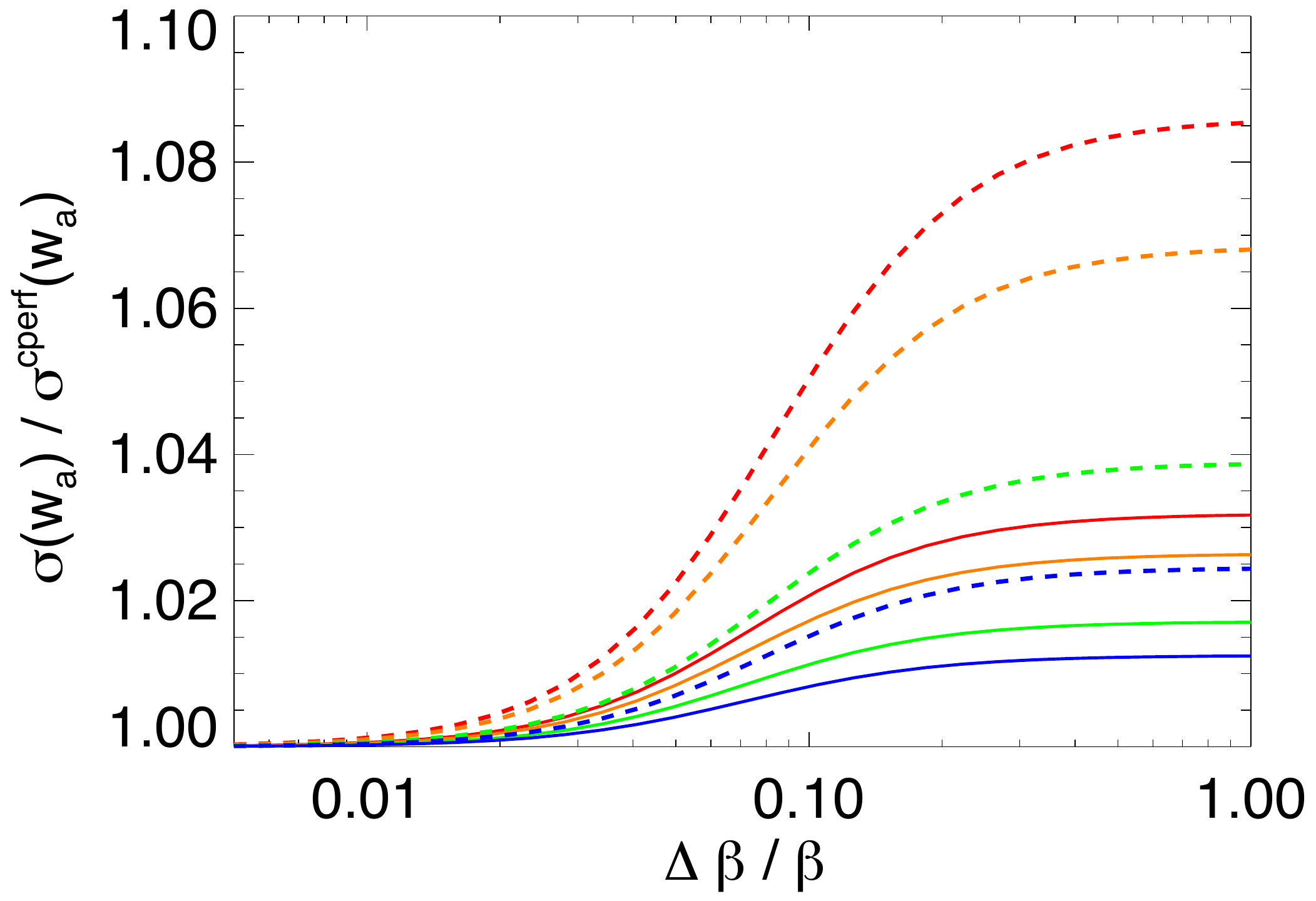}
\caption{
Plot of the degradation in dark energy constraints as a function of  
priors on halo concentration parameters, defined by $c(m,z)=c_{0}[m/m_{*,0}]^{\alpha}(1+z)^{\beta}.$  
Degradation of the constraints on $\wzero$ are shown in the {\em left} panels, while degradation of 
$\wa$ constraints are shown in the {\em right} panels.  The {\em top} panels show degradation 
as a function of the fractional prior uncertainty in the normalization of the mass 
concentration relation $\czero$.  The {\em middle} row of panels shows 
degradation as a function of the power law index describing the dependence of concentration 
on mass, $\alpha$, and the {\em bottom} panels show the degradation as a function of the power-law index 
describing the dependence of concentration on redshift, $\beta$.
Different levels of priors on photo-z parameters 
are color-coded as labeled in the middle panels.  
}
\label{fig:wcons_conc}
\end{figure*}
%-----------------------------------------------------------------------------------------------------

In Figure~\ref{fig:wcons_conc} we display the degradation in the statistical constraints on $\wzero$ (left panels) 
and $\wa$ (right panels) due to uncertainty in the halo concentration parameters $\czero$, $\alpha$, and $\beta$ [Eq.~(\ref{eq:cmz})].  
The precision level of the calibration of $\czero$ appears as the horizontal axis in the top panels, 
$\alpha$ in the middle panels, and $\beta$ the bottom panels.  Along each row of panels in Fig.~\ref{fig:wcons_conc}, 
only the parameter labeled on the horizontal axes has uncertain prior knowledge.  In other words, the remaining two concentration 
parameters are treated as perfectly known for simplicity.  As in Fig.~\ref{fig:wcons_dpmati}, each curve is normalized to the constraint 
that would be realized in the limit of perfect knowledge of the halo structure parameters, $\sigma^{\mathrm{cperf}}$. Note however, 
that this baseline constraint has been recomputed for each assumed value of photometric redshift uncertainty, as specified by the 
color-coding of each curve, so that the degradation represents only that amount of additional degradation due to uncertainty about 
halo structure.  Thus for each level of photometric calibration precision appearing in the legend, $\nspec = 5000,\,10^{4},\, 10^{5},$~and~$10^{6},$ 
the vertical axis value gives the degradation in the $\wzero (\wa)$ constraints strictly due to uncertainty in the parameter labeled on the horizontal axis.

In all cases, the degradation of dark energy parameters is relatively modest.  In particular, the degradation induced by 
halo structure uncertainty alone is $\lesssim 15\%$ for all reasonable models, and significantly less if photometric 
redshifts are well calibrated or if galaxy clustering statistics are employed. 
 This is in qualitative agreement with Ref.~\cite{zentner_etal08}, who studied 
calibrating halo structure parameters as a means to account for the influence of baryonic processes on 
the lensing power spectrum.  It is not expected that halo structure alone can account for all of the effects of 
baryonic processes on the power spectrum (e.g., Refs.~\cite{zhan_knox04,rudd_etal08,zentner_etal08,semboloni_etal11}).  
However, combining the results of Fig.~\ref{fig:wcons_dpmati} with Fig.~\ref{fig:wcons_conc} suggests that if 
uncertainty in the power spectrum due to baryonic processes can be modeled by concentrations with a residual of 
order $\sim 1\%$, then dark energy parameter degradation induced by such uncertainty may be limited to quite 
modest values.  It may even be possible to incorporate additional parameters to describe, for example, 
the hot gas components of groups and clusters \cite{zhan_knox04,rudd_etal08,semboloni_etal11}, at 
a relatively modest statistical cost.

%----------------------------------------------------------------------------------------------------------------------------------------------
\subsection{Systematic Errors on Dark Energy}
\label{subsection:systematics}

In this section we explore the related problem of systematic errors on the dark energy equation of state 
parameters induced by uncertainty in predictions of the matter power spectrum.  
To this point, we have already addressed degradation in dark energy parameters induced by 
treating uncertainty in the matter power spectrum as a statistical uncertainty.  In such a calculation, 
the underlying assumption is that a model for the power spectrum is accurate but the parameters of the model 
 are known with imperfect precision. Our aim in this section is to 
elucidate dark energy equation of state errors in the related circumstance of 
a systematic error on power spectrum predictions.  The underlying framework 
is that a model for the power spectrum exists and is assumed to be correct (or at 
least, that it contains the true power spectrum within its parameter set), but 
the model misestimates the power spectrum over some range of wavenumbers.  
This situation may be the most relevant to forthcoming data analyses if, as a specific example, 
the dominant errors in power spectrum predictions stem from systematic errors 
in the numerical treatment of baryons.  The result of parameter inferences 
that use such systematically-offset theoretical power spectra will be systematically-offset 
dark energy equation of state estimators.  In this section, we quantify the systematic errors in 
dark energy equation of state parameters induced by systematic offsets in the matter 
power spectrum.  Not surprisingly, the results of the previous section will be a 
useful guide for anticipating and understanding the results of this section.  
As before, we begin with our more general $\dpmati$ model and later 
address systematic errors in the halo-based model.  

%-----------------------------------------------------------------------------------
\subsubsection{The $\dpmati$ Model}
\label{subsubsection:wbias_dpmati}

We begin our treatment of systematic errors with our $\dpmati$ model.  This model treats the predicted power spectrum as a 
sequence of bandpowers distributed evenly in $\log(k)$.  The generality of this model enables us to address 
a specific, important concern, namely: How does the severity of the induced systematic error in the dark energy equation of 
state depend on the comoving scale $k$ at which the prediction for $\pmatk$ is erroneous?  In so doing, we can prescribe how 
well the theoretical matter power spectrum should be calibrated as a function of scale, setting a specific 
goal for large-scale simulation efforts.

%---------------------------------------------------------------------------------------------------
\begin{figure*}[t!]
\centering
\includegraphics[width=11.0cm]{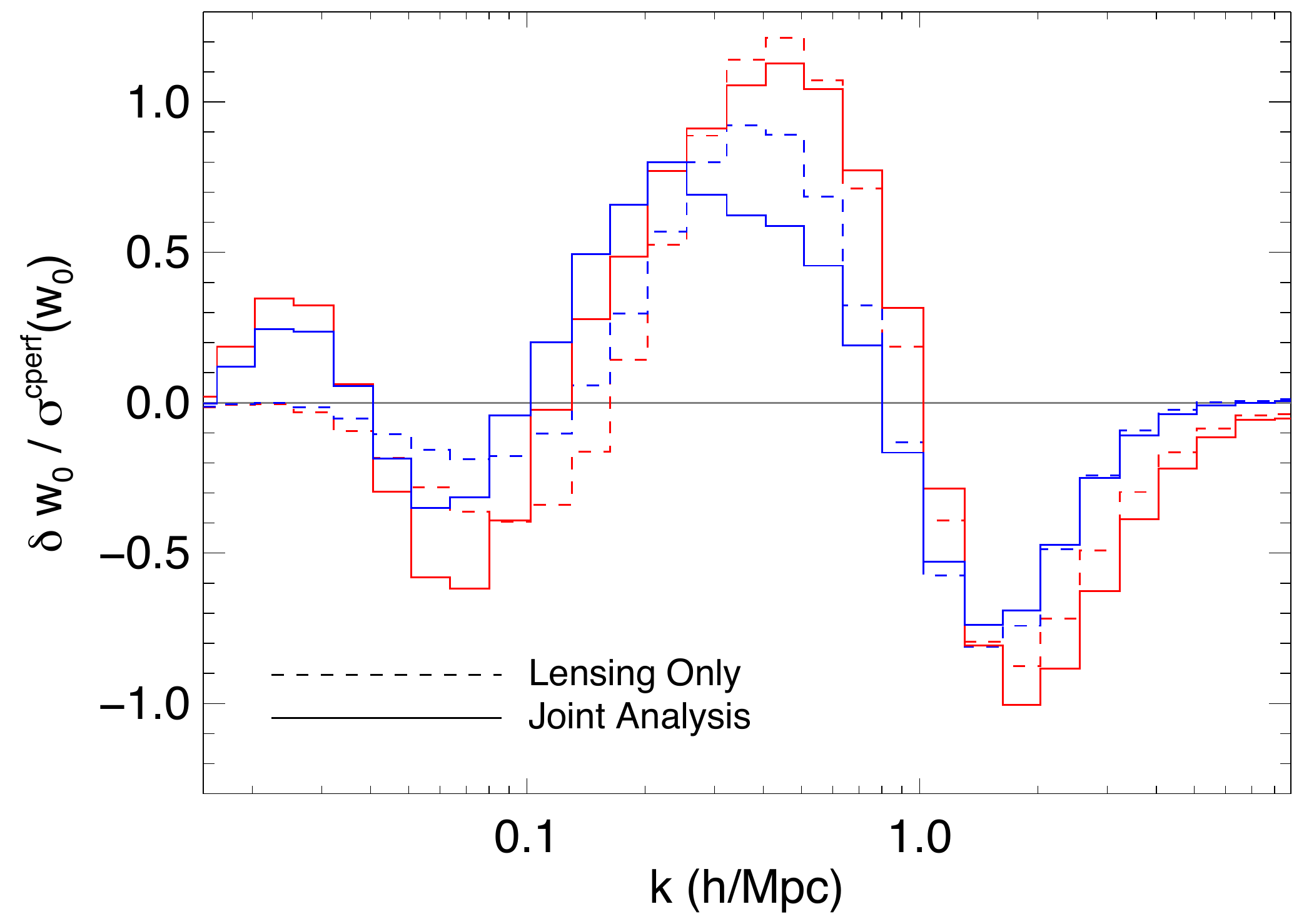}
\includegraphics[width=11.0cm]{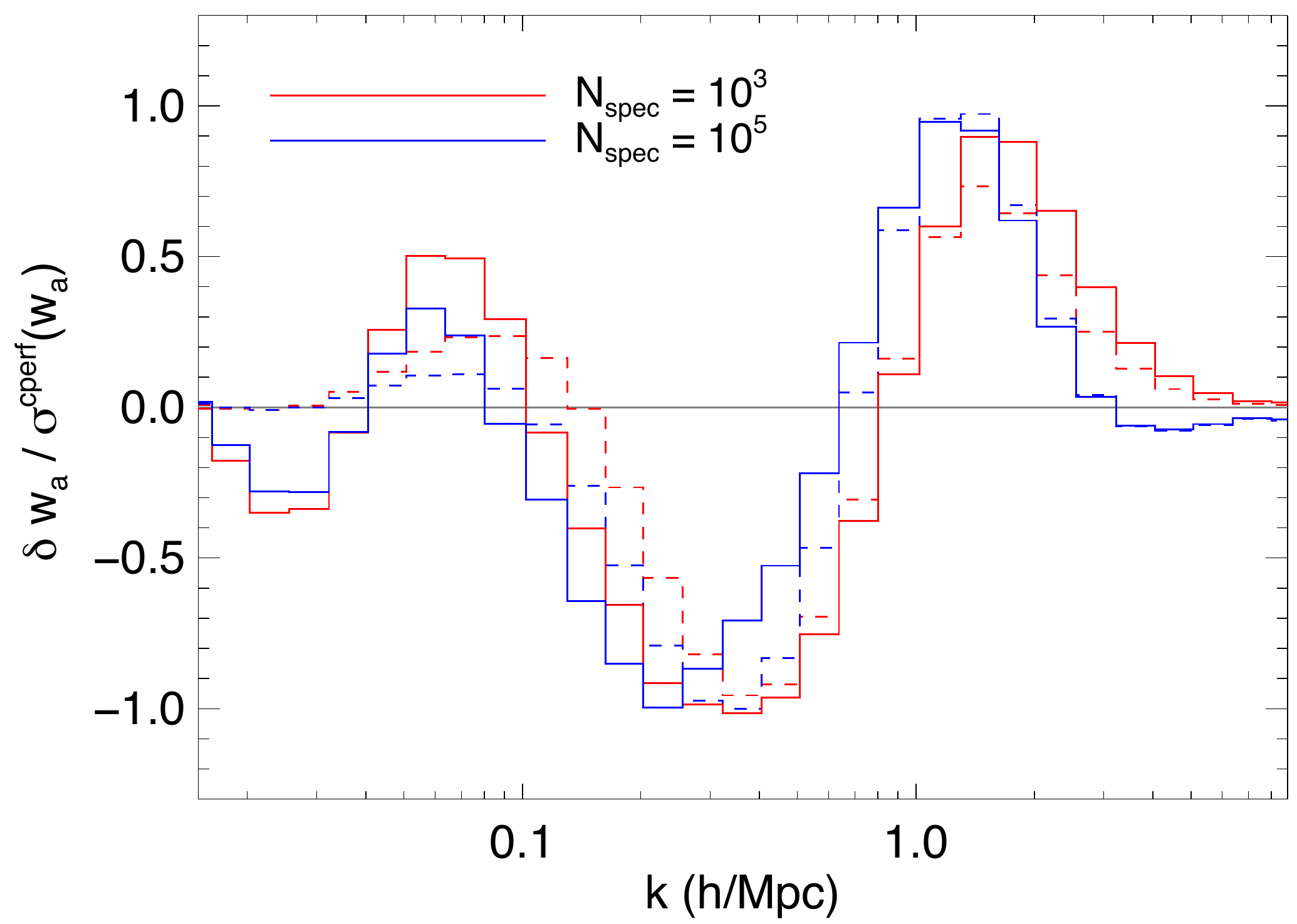}
\caption{
Plot of the systematic error in the inferred value of $\wzero$ (top panel) and $\wa$ (bottom panel) induced by a systematic misestimation of the matter power spectrum $\pmatk$ over a small range of wavenumbers. The value of the induced systematic is scaled in units of the statistical uncertainty of an LSST- or Euclid-like survey, and is plotted against the scale at which $\pmatk$ has been incorrectly predicted by $5\%$ over a range of wavenumbers of width $\Delta k/k=0.1.$ See the text for a detailed description of how these results can be rescaled for $\pmatk$ errors spanning a range of wavenumbers of a different width.
Results pertaining to an analysis that only uses weak lensing information appear as dashed curves, a joint analysis that includes galaxy clustering as solid curves. We show results for two different levels of photo-z calibration, with the red curves corresponding to $\nspec=10^{3},$ and the blue curves to $\nspec=10^{5}.$  
}
\label{fig:biasvk}
\end{figure*}
%-----------------------------------------------------------------------------------------------------

In Fig.~\ref{fig:biasvk} we have plotted the systematic error induced on $\wzero$ (top panel) and $\wa$ (bottom panel) 
by a $5\%$ error in $\pmatk$ at the comoving scale $k$ labeled along the horizontal axis.  
The induced systematic errors have been calculated according to Eq.~\ref{eq:fishersystematic}, 
where the systematic shift to the observables is induced by introducing a $5\%$ systematic error in $\pmatk$ over a range of wavenumbers 
with a width that is $10\%$ of $k_{\mathrm{err}},$ the scale at which the error is centered, viz.
\beq
\label{eq:pmatkpert}
\frac{\Delta\pmatk}{\pmatk} = \left\{
\begin{array}{lr}
0.05 & :  0.95k_{\mathrm{err}} \le k \le 1.05k_{\mathrm{err}} \\
0 & : \mathrm{otherwise}
\end{array}
\right.
\eeq
The curves in Fig.~\ref{fig:biasvk} are color-coded according to the priors on the photo-z distribution, and the magnitude of the 
induced systematic error has been normalized by $\sigma^{\mathrm{cperf}},$ the statistical uncertainty of the 
parameter assuming perfect prior knowledge of each of the $\dpmati$ but with photo-z uncertainty at the level 
given by the color-coding of the curve and the observables included in the analysis. 

Each curve approaches zero at large and small scales. 
Very large scale lensing correlations contribute little constraining power on dark energy, and so it should be expected that scales larger than $k\lesssim10^{-2}$ $\hmpcinv$ should contribute comparably little to the error budget; errors in $\pmatk$ on scales smaller than $k\gtrsim$ $5\hmpcinv$ are comparably tolerable because we make no use of correlation information from multipoles greater than $\ellmax=3000.$ From the wavenumber-range containing the peaks in curves in Fig.~\ref{fig:biasvk} we can see that the most significant dark energy biases induced by systematic errors in $\pmatk$ come from errors on scales $0.05$ $\hmpcinv \lesssim k \lesssim3$ $\hmpcinv.$ This finding is consistent with the results presented in Fig.~\ref{fig:ht1a}, in which we can see that this is the same range of scales over which the tightest statistical constraints can be obtained by self-calibrating the parameters $\dpmati.$ Both of these figures thus illustrate that the weak lensing information about dark energy that will be available to future very-wide-area surveys such as LSST or Euclid comes primarily from gravitational lensing events produced by perturbations on scales $0.05$ $\hmpcinv \lesssim k \lesssim3$ $\hmpcinv,$ in good agreement with previous results \cite{huterer_takada05}. 

Although almost all of the curves are limited to systematic biases $\delta(\wzero,\wa)\lesssim1\sigma(\wzero,\wa),$ we remind the reader that the y-axis value gives the magnitude of the bias when the systematic error in $\pmatk$ is isolated to just a single bin of wavenumbers of width $\Delta k / k =0.1,$ {\em and} when the magnitude of the power spectrum error is $5\%.$ However, we only chose these particular values for the sake of making a definite illustration, and so for the results in Fig.~\ref{fig:biasvk} to be useful to the calibration program it will be necessary to scale the systematics we predict according to the particular details of the matter power spectrum error whose consequences are being estimated. We give several examples of this below to illustrate the utility of our calculations.

Suppose there is a $3\%$ error in $\pmatk$ made over a range of $\Delta k/k=0.4,$ centered at $k\approx2$ $\hmpcinv.$ For definiteness, consider a lensing-only analysis with photo-z uncertainty modeled by $\nspec=10^3.$ The y-axis value of the corresponding (dashed, red) curve at $k\sim2$ $\hmpcinv$ is $\delta\wzero=0.9\sigma(\wzero).$ This value needs to be rescaled by a value of $3/5$ to account for the difference between the magnitude of this example's actual error in $\pmatk$ and the $5\%$ error plotted in Fig.~\ref{fig:biasvk}; additionally, scaling by a factor of $4$ is necessary to account for the fact that the range of scales over which the error is operative spans $4$ of our bins\footnote{Simple, linear scaling is a very good approximation when correcting for the magnitude of the $\pmatk$ error, but this prescription is only approximately correct when rescaling according to the width of the range of scales over which the error is made. We find that the derivatives of lensing power spectra with respect to $\dpmati$ parameters are stable to roughly factor-of-five changes in numerical step-size, and so simple, linear scaling will be appropriate so long as the width of the wavenumber range is less than $\Delta k/k\lesssim0.5.$}, giving an estimate of $\delta\wzero\approx2.2\sigma(\wzero).$ 

If a $3\%$ $\pmatk$ error, again spanning $4$ of our $\Delta k/k=0.1$ bins, is instead made at $k\approx0.8$ $\hmpcinv,$ the estimation method of the first example naively implies that there is zero systematic error associated with such a power spectrum misestimation because the error is centered at a wavenumber at which  $\delta\wzero$ changes sign, and so the contributions to the net dark energy systematic to the left and right of $k\approx0.8$ $\hmpcinv$ appear to cancel. Such cancellations are not necessarily spurious, and in fact a very general formalism for choosing a set of nuisance parameters specifically designed to take advantage of this phenomenon has recently been proposed \cite{norena_etal11}. However, because it may not be known if the sign of the matter power spectrum error inducing the biases also changes sign over the range of wavenumbers on which the error is made, it may not be possible to exploit this sign change to minimize the net effect of the error. In such a case, we advocate assuming the worst-case scenario, that biases produced by errors in multiple bins in wavenumber conspire to contribute additively; by construction this will yield a conservative estimate for the systematic induced by the error in $\pmatk.$ Thus for this example, the absolute value of the y-axis values should be used to estimate the net dark energy systematic. The magnitude of the dashed, red curve for this particular systematic peaks at $\delta\wzero=0.5\sigma(\wzero)$ at the endpoints of its operative range, $0.6$ $\hmpcinv\lesssim k\lesssim1$ $\hmpcinv;$ the curve approaches these maxima from its value of zero at $k=0.8$ $\hmpcinv$ and so we approximate this as an $0.25\sigma(\wzero)$ error spanning $4$ of our bins. Thus for a $3\%$ systematic error in $\pmatk$ spanning the range $0.6$ $\hmpcinv\lesssim k\lesssim1$ $\hmpcinv,$ our final estimate is given by $\delta\wzero=\left[0.25\times4\times(3/5)\right]\sigma(\wzero)=0.6\sigma(\wzero).$ 

For power spectrum errors made over very broad ranges of wavenumber, the simple linear scaling of $\Delta k / k$ is no longer appropriate and one must rely on the full machinery of our calculation to integrate the absolute value of $\delta\wzero / \sigma(\wzero)$ over the scales over which the error is operative. This is also useful to conservatively estimate an ultimate target goal for the matter power spectrum calibration effort. When performing this integration on scales $1 \hmpcinv \lesssim k \lesssim 5 \hmpcinv,$ the worst-case estimate is a systematic error of $\delta(\wzero)\approx3-4\sigma(\wzero);$\footnote{The exact number depends on the level of photo-z calibration as well as the choice of observables.} if one instead integrates a $5\%$ $\pmatk$ error over the entire range of wavenumbers $0.01 \hmpcinv\lesssim k\lesssim 5\hmpcinv,$ a worst-case estimate of the coherently contributing biases ranges from $9-13\sigma.$ To ensure that dark energy systematics are kept at or below the level of statistical constraints, $0.5\%$ accuracy in the prediction for $\pmatk$ over the entire range of $0.01 \hmpcinv\lesssim k\lesssim 5\hmpcinv$ will be required of the simulations calibrating the matter power spectrum. Note that these requirements are somewhat more restrictive than estimations from previous work \cite{huterer_takada05}.

%-----------------------------------------------------------------------------

\subsubsection{Dependence on Multipole Range}
\label{subsubsection:ellmax}

\begin{figure*}[t!]
\centering
\includegraphics[width=12.0cm]{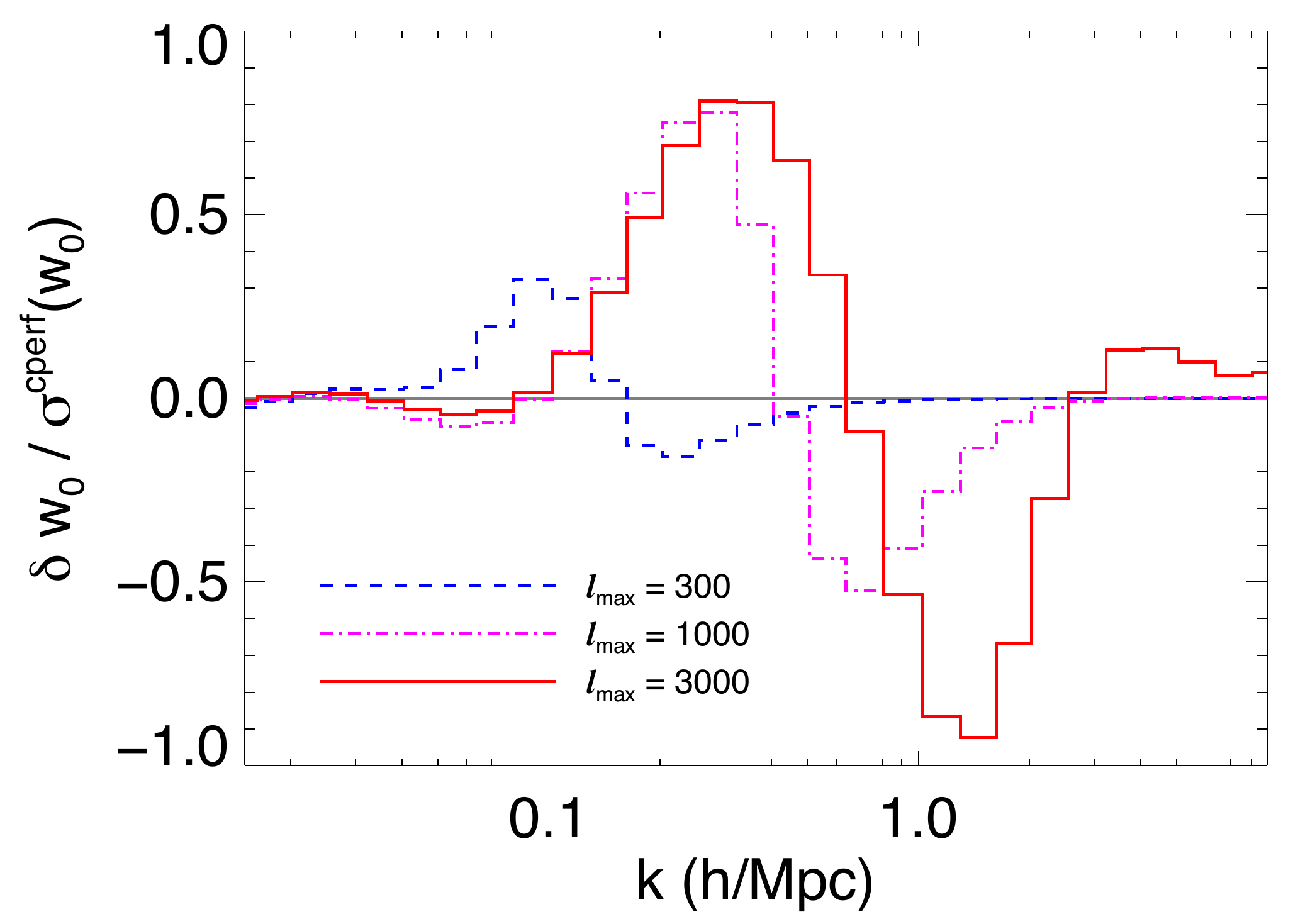}
\caption{
Plot of the systematic error in the inferred value of $\wzero$ (top panel) induced by a systematic misestimation of the matter power spectrum $\pmatk.$ This plot is identical to Figure 5 except the curves illustrate weak lensing-only results in the limit of perfect core calibration ($\nspec\rightarrow\infty$) but with different choices for the maximum multipole. Naturally, the sensitivity of the experiment to small-scale $\pmatk$ errors decreases as $\ellmax$ decreases.
}
\label{fig:bias_v_ellmax}
\end{figure*}

The results presented above in~\S\ref{subsubsection:wbias_dpmati} depend sensitively on $\ellmax,$ the maximum multipole used in the cosmic shear analysis. Naturally, as $\ellmax$ increases the matter power spectrum must be modeled with greater precision and to smaller scales. Figure \ref{fig:bias_v_ellmax} represents a simple illustration of this point. Each curve in Fig.~\ref{fig:bias_v_ellmax} pertains to a cosmic shear-only experiment in the limit of perfect knowledge of the photo-z distribution, but with different choices for $\ellmax$ color-coded according to the legend. The axes are the same as those in the top panel of Fig.~\ref{fig:biasvk}, and so this plot shows how $\wzero$ biases induced by $\pmatk$ errors change with the choice for the maximum multipole used in the lensing analysis.

%---------------------------------------------------------------------------------------------------
%-----------------------------------------------------------------------------------------------------

At the end of ~\S\ref{subsubsection:wbias_dpmati} we described how to use results such as those appearing in Fig.~\ref{fig:bias_v_ellmax} to estimate the precision to which the matter power spectrum must be predicted in order to guarantee that dark energy biases induced by $\pmatk$ systematics are kept at or below the level of the statistical constraints. Briefly, one assumes the worst case scenario, that the sign of the $\pmatk$ errors conspire to contribute coherently to the dark energy bias. In this case, to estimate the most severe dark energy bias that could be induced by an error in $\pmatk$ made over a range of wavenumbers, one simply adds the absolute value of the relevant curve over the relevant range of wavenumbers. To obtain a more optimistic estimation, one could suppose that the errors in each bin are perfectly uncorrelated, in which case they may be treated as independent Gaussian random variables so that their net contribution to the error budget is computed by adding the individual contributions in quadrature. In either the pessimistic or optimistic case, one obtains a precision requirement by finding the magnitude of the $\pmatk$ error that would result in the sum described above equal to unity, since this would imply that the net systematic bias on the dark energy parameter is equal to the statistical constraint on that parameter (recall that the curves in Figures \ref{fig:biasvk} and \ref{fig:bias_v_ellmax} plot the systematics in units of the statistical uncertainty of the survey). Ideally, of course, the goal of the $\pmatk$ calibration program is to achieve sufficient precision such that the systematics are well below the level of statistical constraints, and so setting this sum to unity simply provides a guideline for the calibration.

%---------------------------------------------------------------------------------------------------
\begin{figure*}[t!]
\centering
\includegraphics[width=15.0cm]{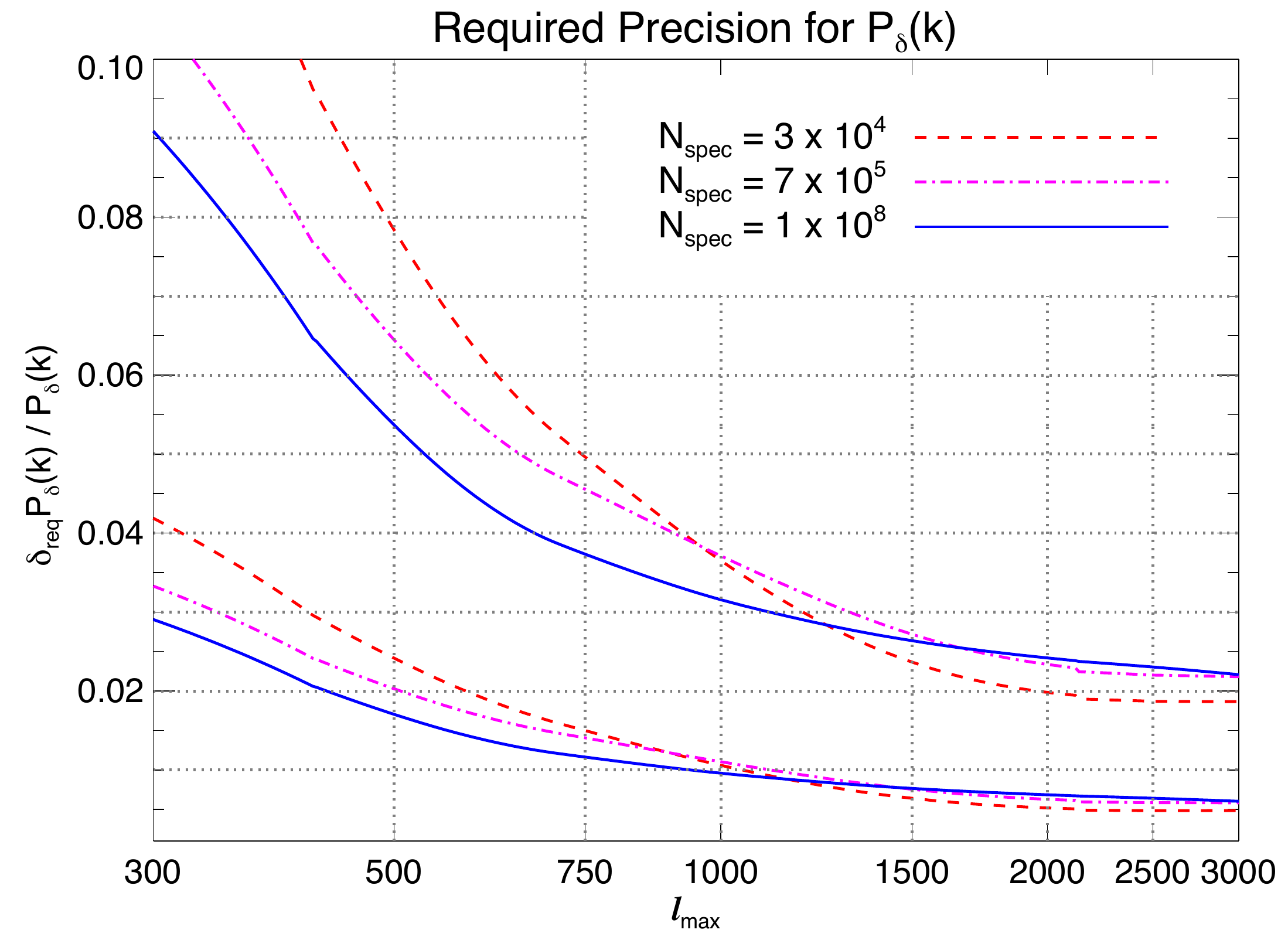}
\caption{
Plot of the required precision with which the matter power spectrum must be predicted as a function of $\ellmax.$ The top curves represent an optimistic estimate, the bottom curves a pessimistic estimate. All curves represent results for a cosmic shear-only analysis; calculations for different levels of photo-z precision are color-coded according to the legend. To guarantee that matter power spectrum errors do not contribute significantly to the dark energy error budget, the precision in the prediction for $\pmatk$ must reach the level illustrated by the bottom curves. One  uses the results in Figure 8 to estimate the wavenumber to which the level of precision plotted here must be attained. See text for details concerning the calculation of these estimates.
}
\label{fig:prec_v_ellmax}
\end{figure*}
%-----------------------------------------------------------------------------------------------------

In Fig.~\ref{fig:prec_v_ellmax} we have performed the calculation of the power spectrum precision requirement as a function of $\ellmax.$ All curves pertain to weak lensing-only experiments\footnote{We have not illustrated the significance of including galaxy clustering statistics because it is a relatively minor effect, as evidenced by Fig.~\ref{fig:biasvk}.}, with the level of photo-z precision coded according to the legend. The top three curves correspond to the optimistic precision estimate (errors in each bin are added in quadrature), the bottom three curves to the pessimistic estimate (the absolute value of the error in each bin are added). The requirements plotted in Fig.~\ref{fig:prec_v_ellmax} are set according to the magnitude of the systematics in $\wzero;$ the $\wa$-based requirements are very similar. 

%---------------------------------------------------------------------------------------------------
\begin{figure*}[t!]
\centering
\includegraphics[width=15.0cm]{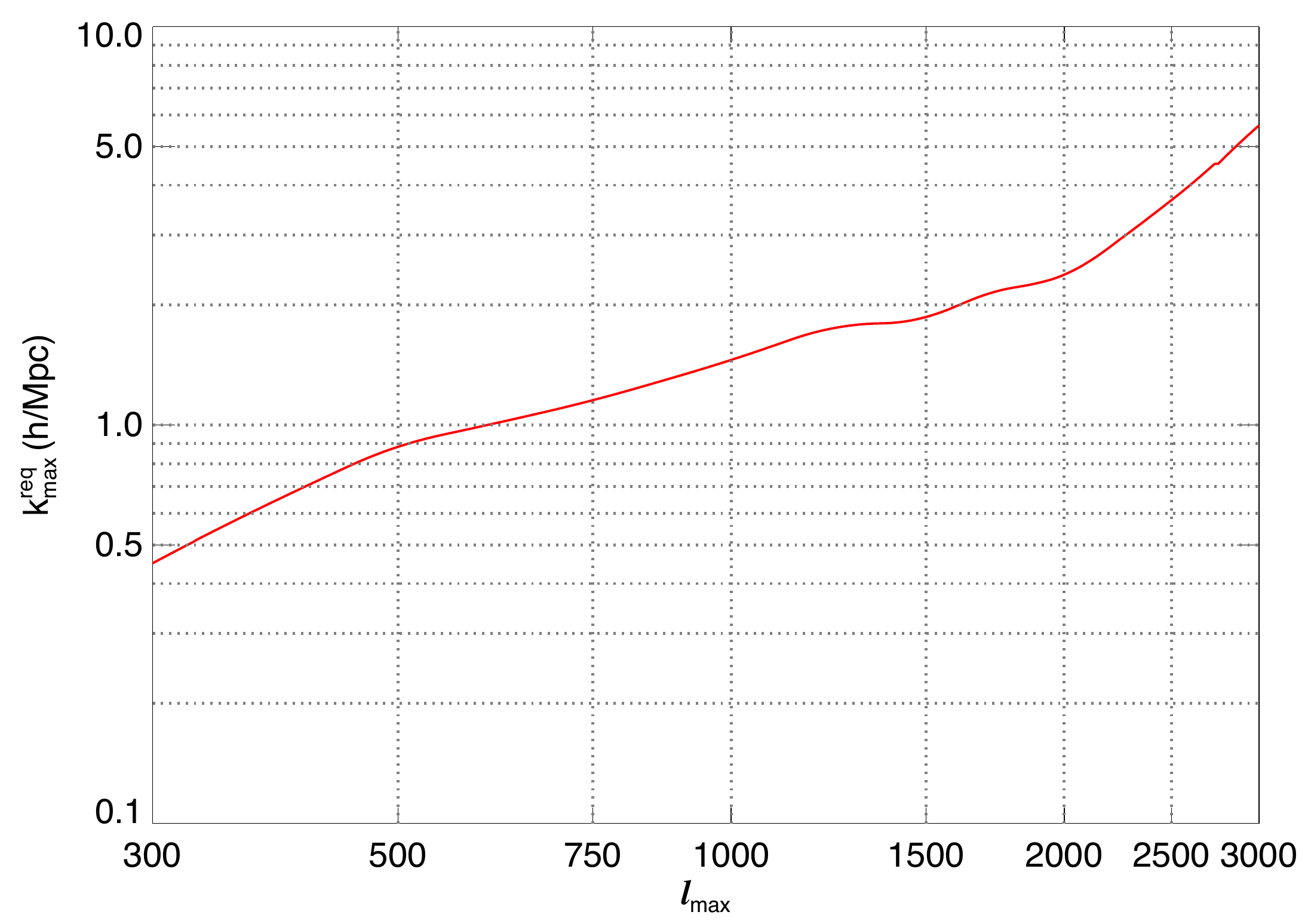}
\caption{
Plot of $\kreq,$ the wavenumber to which the matter power spectrum must be predicted, as a function of $\ellmax.$ The quantity $\kreq$ is insensitive to the level of photo-z precision, and so the same $\kreq$ pertains to all values of $\nspec.$ One uses the results of Figure 7 to estimate the level of precision in the prediction for $\pmatk$ that needs to be attained for all $k<\kreq.$
}
\label{fig:kreq_v_ellmax}
\end{figure*}
%-----------------------------------------------------------------------------------------------------

The optimistic calculation assumes that the $\pmatk$ errors at different wavenumber are completely independent; the pessimistic calculation assumes that the errors are perfectly correlated. Because the level of correlation between matter power spectrum errors at different wavenumbers will not be known, we stress that for a given $\ellmax$ the only way to guarantee that $\pmatk$ errors do not contribute significantly to the dark energy error budget is to attain the level of precision illustrated by the bottom curves.

In Fig.~\ref{fig:prec_v_ellmax}, the crossing of the $\pmatk$ precision requirement curves pertaining to different levels of $\nspec$ may seem somewhat counterintuitive; one might expect that improving photo-z uncertainty can only lead to more stringent demands on the accuracy of the prediction for the matter power spectrum. This intuitive expectation is supported by a theorem proved Appendix A of Ref.~\cite{bernstein_huterer09}, in which the authors demonstrate that the net $\Delta\chi^2$ induced by a systematic error is always reduced by the addition of (unbiased) prior information. However, as shown in Appendix B of the same paper when marginalizing over multiple parameters, $\Delta\chi^2$ {\em per degree of freedom} may increase. In our case, for large maximum multipoles ($\ellmax\gtrsim1000$) adding prior photo-z information and marginalizing over our $62$ photo-z parameters leads to a mild increase in the $\wzero$ bias, a fact which we have traced to a mild difference in the degeneracy between $\wzero$ and $\Omega_{\Lambda}$ for different values of $\nspec.$

From Fig.~\ref{fig:bias_v_ellmax} it is evident that for smaller choices of $\ellmax$ dark energy biases are less sensitive to matter power spectrum errors on small scales. For each $\ellmax$ there is a maximum wavenumber, $\kreq,$ such that systematic errors in $\pmatk$ for $k>\kreq$ do not produce significant biases in dark energy parameters. We estimate $\kreq$ by finding the bin in wavenumber at which the induced $\wzero$ bias becomes less than $10\%$ of the maximum magnitude that $\delta\wzero$ attains in any bin. Because of the steepness of the scaling of $\delta(\wzero) / \sigma(\wzero)$ with $k$ on scales smaller than the wavenumber at which the systematics attain their maximum magnitude, we find that our $\kreq$ estimations are insensitive to the choice for this percentage. We present our results for $\kreq$ as a function of $\ellmax$ in Fig.~\ref{fig:kreq_v_ellmax}.

The results in this section provide a set of concrete benchmarks for the campaign of numerical simulations designed to calibrate the prediction for the matter power spectrum, as well as a guideline for choosing the maximum multipole that should be included in any cosmic shear analysis. For a given $\ellmax,$ one uses the results presented in Fig.~\ref{fig:prec_v_ellmax} to estimate the precision with which $\pmatk$ must be predicted on all scales $k<\kreq,$ where the $\kreq$ estimate appears in Fig.~\ref{fig:kreq_v_ellmax}. Of course choosing smaller values of $\ellmax$ naturally decreases the constraining power of the survey, and so the results we present here can be used to inform the optimal choice for $\ellmax$ that balances the need for statistical precision against the threat of matter power spectrum systematics.

%-----------------------------------------------------------------------------

\subsubsection{Halo Model}
\label{subsubsection:wbias_conc}

\begin{figure*}[t!]
\centering
\includegraphics[width=10.0cm]{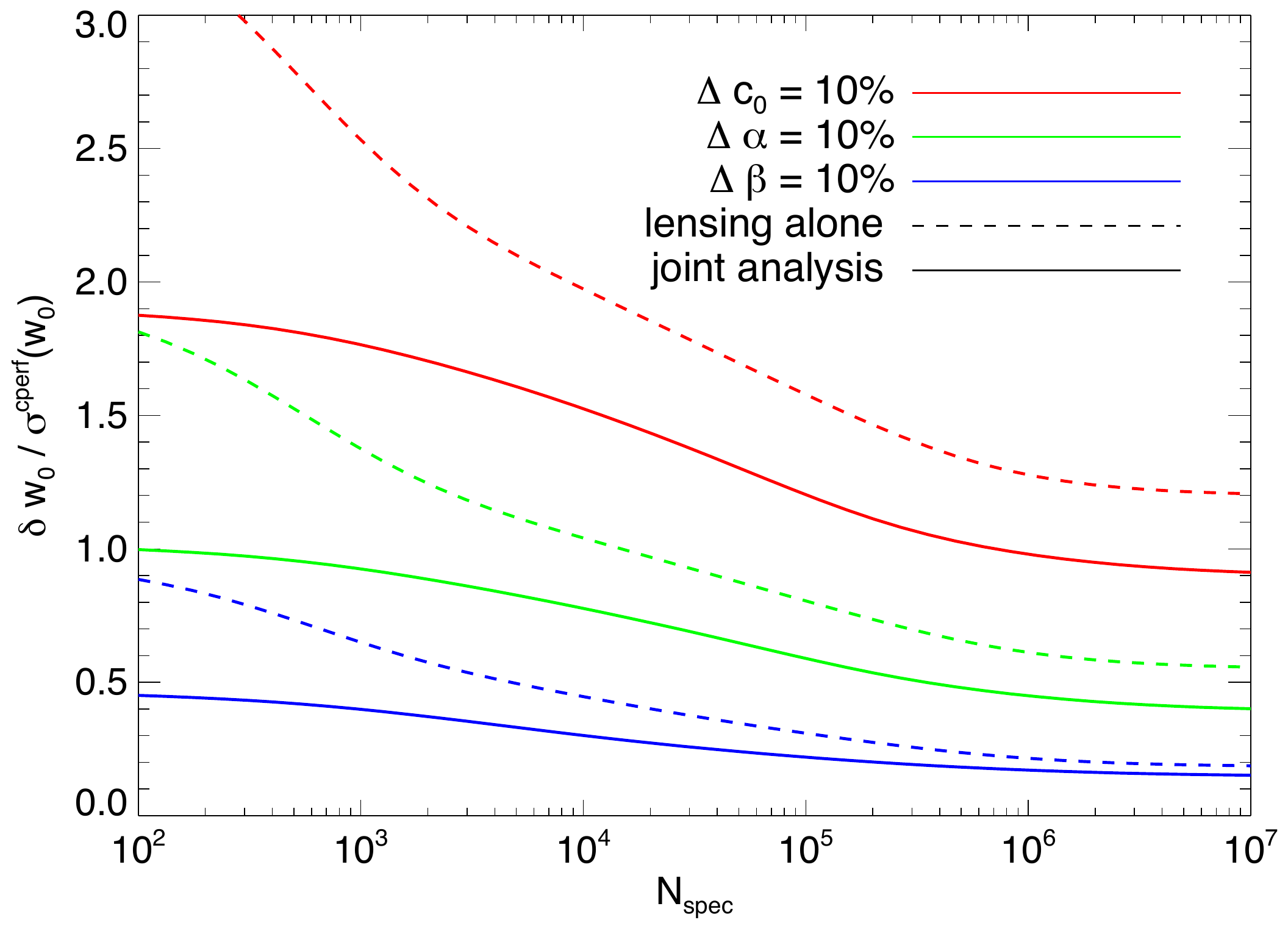}
\includegraphics[width=10.0cm]{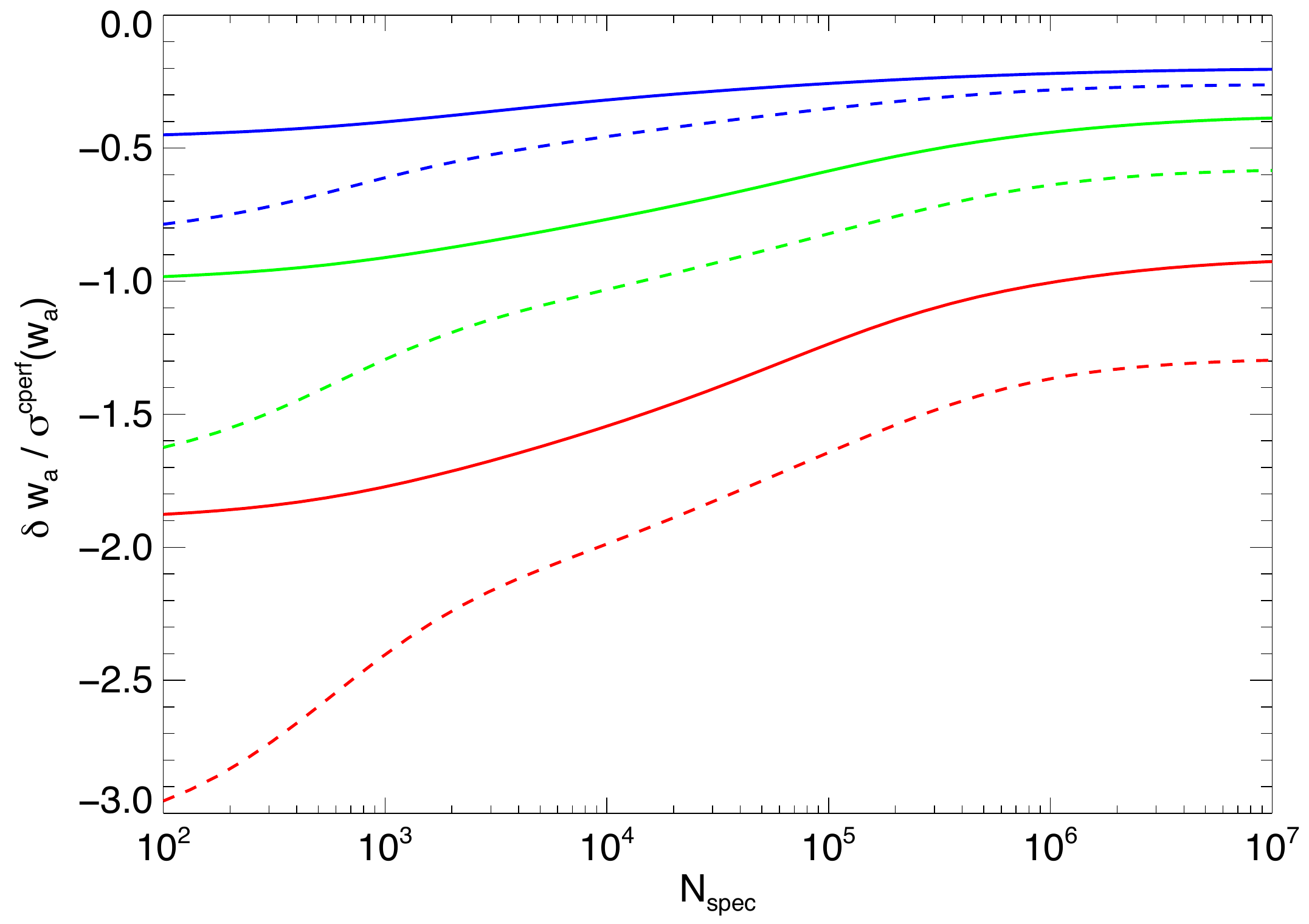}
\caption{
In the {\em top} panel we plot the systematic error on $\wzero$ against $\nspec,$ the quantity encoding the precision with which the photo-z distribution has been calibrated. The same quantity is plotted for $\wa$ in the {\em bottom} panel. Dark energy systematics $\delta(\wzero,\wa)$ have been scaled by $\sigma^{\mathrm{cperf}}(\wzero,\wa),$ the statistical constraints on the parameter at the level of $\nspec$ given by the horizontal axis value of $\nspec.$ The systematic errors are induced by $10\%$ misestimations of our three halo concentration parameters $c_0,$  $\alpha,$ and $\beta.$ Results for each parameter are color-coded according to the legend in the {\em top} panel.
The concentration parameters are defined by $c(m,z)=c_{0}[m/m_{*,0}]^{\alpha}(1+z)^{\beta}.$ Dashed curves correspond to results for an analysis using weak lensing information only, solid curves an analysis employing galaxy clustering information together with weak lensing in a joint analysis.
}
\label{fig:wbias_conc_v_nspec}
\end{figure*}

In Fig.~\ref{fig:wbias_conc_v_nspec} we illustrate our results for the propagation of 
systematic errors in halo concentration parameters through to $\wzero$ (top panel) and $\wa$ (bottom panel).  
Each curve corresponds to a calculation in which a single concentration parameter, either $\czero$ (red), 
$\alpha$ (green), or $\beta$ (blue), is systematically offset upwards of its fiducial value by $10\%,$ while assuming that 
all of the concentration parameters are known with perfect accuracy and precision.  
The value of the induced systematic error on $\wzero (\wa)$ has been normalized by the statistical constraints on the parameter at the 
level of photo-z calibration specified by the horizontal axis value for $\nspec.$ We propagate systematic 
errors via Eq.~\ref{eq:fishersystematic}, as in \S~\ref{subsubsection:wbias_dpmati}.

The magnitude of the systematic error induced on dark energy parameters monotonically decreases as $\nspec$ increases. The physical interpretation of this trend applies to nearly all of the results presented in this manuscript, and so we discuss it in detail in~\S\ref{section:discussion}. Briefly, as photo-z priors are relaxed the cosmological interpretation of the weak lensing signal must rely more heavily on precise knowledge of the matter distribution. In the context of the halo model this implies that errors in halo concentrations have more drastic consequences for dark energy parameter inference at lower values of $\nspec.$  

%---------------------------------------------------------------------------------------------------
%-----------------------------------------------------------------------------------------------------

Dark energy systematics are also less severe when galaxy clustering information is included; for each halo parameter, and at every level of photo-z calibration, the solid curves are smaller in magnitude than the dashed. This trend is to be expected as its analogues have manifested in previous sections: the additional information available in galaxy clustering statistics mitigates the consequences for cosmology of errors in the matter power spectrum prediction (see Ref.~\cite{zhan06} for the analogous benefit of mitigating photo-z systematics by including galaxy correlations). Notice that there is a proportionally greater mitigation of the systematics at lower values of $\nspec;$ the interpretation of this observation is somewhat subtle.  When photo-z priors are weak, galaxy clustering information plays a more important role in the self-calibration of the photo-z parameters\footnote{This is an important observation in itself and has been noted elsewhere in the literature (for example, Ref.~\cite{zhan06}). Most of the so-called ``complementarity" of weak lensing and galaxy clustering stems from these signals calibrating each other's nuisance parameters.}, $\sigma^{i}_{z}$ and $z^{i}_{\mathrm{bias}}.$ As discussed in the preceding paragraph, systematic errors induced by incorrect predictions of halo concentrations are more severe when information about the photo-z distribution is limited. Therefore, because including galaxy clustering information at low $\nspec$ has a greater impact on the statistical constraints than when $\nspec$ is very large, there is a concomitantly greater mitigation of the induced systematics associated with halo concentration errors at lower values of $\nspec.$ 

In the limit of very large $\nspec,$ when uncertainty in the photo-z distribution can be neglected, a $10\%$ error in the mean halo concentration $c_0$ induces a systematic error on dark energy parameters that is comparable to or worse than the statistical constraints. This estimation is consistent with previous results (Refs.~\cite{zentner_etal08,hearin_etal09}). For halo concentration errors made at other levels, we can estimate the induced systematic error on the inferred value of dark energy parameters via simple linear scaling\footnote{The stability of the derivatives of our observables with respect to halo concentration parameters over a broad range of numerical step sizes ensures the accuracy of this simple linear scaling.}. For example, as can be seen in Fig.~\ref{fig:wbias_conc_v_nspec}, in the limit of very large $\nspec$ a $10\%$ misestimation of the parameter $c_0$ induces a systematic error $\delta\wzero=1.2\sigma(\wzero)$ for a weak lensing-only analysis; 
thus a $25\%$ error on $c_0$ induces a $\delta\wzero=\left[1.2\times(0.25/0.1)\right]\sigma(\wzero)=3\sigma(\wzero)$ systematic. For the sake of concreteness we conclude this section with the following rough guideline that is based on Fig.~\ref{fig:wbias_conc_v_nspec} evaluated at $\nspec\approx10^{4}$ for a joint analysis: in order to guarantee that dark energy systematics induced by halo concentration errors are kept at or below the statistical constraints, the parameter $c_0$ must be calibrated to an accuracy of $5\%$ or better, the parameter $\alpha$ to an accuracy of $12\%$ or better, and $\beta$ to better than $25\%.$

\section{Discussion}
\label{section:discussion}
%----------------------------------------------------------------------------

We have studied the significance of matter power spectrum uncertainty for weak lensing measurements of dark energy. Our results can serve as an updated guideline for the calibration requirements on theoretical predictions of $\pmatk$ and photometric redshift distributions. The photo-z requirements revise those in Ref.~\cite{ma_etal06}, who modeled nonlinear evolution using the Peacock \& Dodds fitting formula, and were thus overly pessimistic about photo-z calibration. We also revise the requirements for precision in the prediction of $\pmatk$ outlined in Ref.~\cite{huterer_takada05}, who assumed perfect knowledge of the distribution of photometric redshifts and were thus overly optimistic.

Both of our models for uncertainty in the theoretical prediction for the matter power spectrum have been studied previously. Our second model, in which we allow the value of $\pmatk$ to vary freely about its fiducial values in ten bins of bandpower (see~\S\ref{subsection:nlcompare} for a detailed description) is based on the treatment in Ref.~\cite{huterer_takada05}. Our results are in good agreement with theirs, where applicable. We have generalized their results by 1) studying the self-calibration limit of $\pmatk$ uncertainty, 2) including galaxy clustering statistics in the set of observables, and 3) by taking into account uncertainty in the distribution of photometric redshifts. Our motivation to treat uncertainty in $\pmatk$ and the photo-z distribution simultaneously comes from results presented in the Appendix of Ref.~\cite{hearin_etal10}, where the authors showed that the photo-z calibration requirements vary significantly depending on the assumed fiducial model of the $\pmatk$ in the nonlinear regime. This result suggests a nontrivial interplay between the photo-z and matter power spectrum calibration demands. 

The cause of this interplay has a simple physical interpretation. In weak lensing, there is a degeneracy between the redshift of a galaxy whose image is distorted and the typical size of the overdensity responsible for most of its lensing: at fixed angular scale, correlations in the image distortions of sources at high redshift are produced (on average) by overdensities that are larger in comoving size than those producing correlations at low-redshift. The more precisely the photo-z distribution is known, the narrower the range of possible wavenumbers contributing to the lensing signal.  As priors on photo-z parameters are relaxed, the redshifts of the sources are known with decreasing precision, and so more information about the power spectrum is required in order to compensate. Thus at lower values of $\nspec,$ dark energy parameter inference at a fixed level of statistical uncertainty requires more precise knowledge of the matter power spectrum. 

Consequences of this basic physical picture appear throughout this manuscript. For example, in Figures \ref{fig:biasvk} and \ref{fig:wbias_conc_v_nspec}, appearing in~\S\ref{subsubsection:wbias_dpmati} and~\S\ref{subsubsection:wbias_conc}, respectively, the degeneracy between source redshift and length scale manifests as dark energy biases being more severe for the lower values of $\nspec.$ Another example appears in Figure \ref{fig:wcons_conc} of~\S\ref{subsubsection:decons_hm}, in which we can see that the statistical constraints on $\wzero$ and $\wa$ degrade more rapidly with uncertainty in halo concentrations when $\nspec$ is small relative to larger values of $\nspec.$ Similarly, the statistical constraints on dark energy discussed in~\S\ref{subsubsection:decons_dpmati} degrade more rapidly as priors are relaxed on $\dpmati$ at lower values of $\nspec.$ The nontrivial relationship between uncertainty in $\pmatk$ and in the distribution of photometric redshifts clearly illustrates that a detailed and accurate study of the calibration requirements on future imaging surveys requires the simultaneous account of these contributions to the dark energy error budget that we present here.

The significance of uncertainty in halo concentrations for the dark energy program has also been studied previously~\cite{zentner_etal08,hearin_etal09}. Again, we have generalized their calculations and our conclusions are in good agreement with their results, where commensurable. In particular, Ref.~\cite{zentner_etal08} studied the prospects for future weak lensing surveys to self-calibrate halo concentration parameters while simultaneously constraining dark energy parameters. Even with only very modest prior information on photo-z parameters, we agree with the conclusion in Ref.~\cite{zentner_etal08} that the prospect for future imaging surveys to self-calibrate uncertainty in halo concentrations is very promising, especially when galaxy correlation statistics are employed in a joint analysis: the statistical degradation on $\wzero$ from self-calibrating the mean, redshift-zero halo concentration $c_0$ is less than $6\%$ for $\nspec=5000$ (corresponding roughly to $\Delta\sigma_z / \sigma_z \approx 10^{-2}$). The degradation in the constraints is even milder when the photo-z distribution is more precisely characterized. However, this result is provisional in that it relies on halo concentration being the most significant mode in which the power spectrum is uncertain.

Our halo model-based treatment of $\pmatk$ uncertainty is well-motivated by Ref.~\cite{rudd_etal08}, who carried out a suite of numerical cosmological simulations including hydrodynamics with a variety of energy feedback mechanisms. One of the salient conclusions of Ref.~\cite{rudd_etal08} is that the effects of baryonic physics on the matter power spectrum can be well-modeled as an enhancement to the mean concentration of dark matter halos. Thus we chose our first model of uncertainty in $\pmatk$ with the intention to study the requirements of the dark energy program for precision in our ability to predict the effects of baryons on the large-scale distribution of matter. However, recent results from the OverWhelmingly Large Simulation (OWLS) project \cite{schaye_etal10} suggest that an energy feedback mechanism modeling the effects of AGN is necessary to reproduce the characteristics of groups of galaxies \cite{mccarthy_etal10}.  In a recent study based on these results \cite{semboloni_etal11}, the authors found that a multicomponent halo model with a gas profile that is independent from the dark matter profile can accurately model the power spectra in the OWLS project. We note, however, that not even N-body simulations have achieved the desired precision ($<0.5\%$) over the full range $0.1$ $\hmpcinv\lesssim k\lesssim5$ $\hmpcinv$ required of the $\pmatk$ calibration \cite{heitmann_etal09}. Nonetheless, these results are intriguing and suggest that a more complicated model than the one we consider here may be necessary to fully encapsulate the baryonic modifications to the matter power spectrum. We leave the development and exploration of such a model as a task for future work.

When the assumption that the halo model accurately characterizes all the gross features of $\pmatk$ is relaxed, self-calibrating the matter power spectrum is very likely to be infeasible. To see this, we turn back to our second, more conservative model of matter power spectrum uncertainty. As shown in Fig.~\ref{fig:ht1a}, even in the limit of perfectly precise prior knowledge on the photo-z distribution, future surveys are unable to self-calibrate the value of $\pmatk$ to better than $7\%$ on any scale, rendering $\pmatk$ as a dominant component in the error budget and increasing errors on $\wzero$ and $\wa$ by a factor of $3$ or more. As discussed in~\S\ref{subsubsection:wbias_dpmati}, ensuring that dark energy systematics induced by misestimations of $\pmatk$ are kept at or below the level of statistical uncertainty of an LSST- or Euclid-like survey, the theoretical prediction for the matter power spectrum will need to be accurate to at least $0.5\%$ or better on all scales $k\lesssim5$ $\hmpcinv.$ These results reinforce the necessity of an aggressive campaign of numerical cosmological simulations if surveys such as LSST or Euclid are to achieve their potential as dark energy experiments.

To contextualize these findings with the current state-of-the-art in numerical simulations, we first compare this requirement to the results from the Coyote Universe project \cite{heitmann_etal08b}, a suite of nearly 1,000 N-body (gravity only) simulations spanning $38$ fiducial $w\mathrm{CDM}$ cosmologies. To date, this is the most ambitious campaign of N-body simulations yet performed with the aim to robustly calibrate $\pmatk$ over the full range of scales relevant to weak lensing. The power spectrum emulator based on their results has recently been completed \cite{lawrence_etal10}, and in Ref.~\cite{heitmann_etal09} the authors demonstrate that results from their simulations can be used to model $\pmatk$ with sub-percent accuracy on all scales $k\lesssim1$ $\hmpcinv.$ In addition to the need to expand the range of scales over which this level of precision has been attained, the Coyote Universe does not account for the nonlinear effects of neutrino mass~\cite{abazajian_etal05}, which have been recently established~\cite{bird_etal11} to introduce percent-level changes to the standard Smith et al.~method of prediction. Moreover, Coyote's $1\%$ precision only applies to $w\mathrm{CDM}$ cosmologies; extending this level of precision to dynamical dark energy models is an active area of current research on in this field, for example, Refs.~\cite{casarini_etal11b,mcdonald_etal06,rasera_etal10,joudaki_etal09}. 
These complications aside, the Coyote Universe project is, by itself, insufficient to completely calibrate the matter power spectrum because N-body simulations neglect the effect that baryonic gas has on $\pmatk.$ Suites of hydrodynamical simulations such as the OWLS project \cite{vandaalen_etal11} discussed above will be essential contributions to the calibration program. Continued improvement both in N-body and hydrodynamical simulations will clearly be necessary in order to meet the calibration requirements we present here. 

\subsection{Caveats}
\label{subsection:caveats}

In updating the photo-z precision requirements for future imaging surveys we have quantified the uncertainty in the photo-z distribution in terms of $\nspec.$ We reiterate here an important difference between the meaning of $\nspec$ in our forecasts and elsewhere in the literature. In this work, the quantity $\nspec$ defines a one-parameter family of priors on the photo-z distribution via Eqs.~\ref{eq:photozpriors} \& \ref{eq:photozpriors2}. In practice the actual number of galaxies in the calibration sample will likely need to be larger than $\nspec,$ for example because it will be challenging to obtain a calibration sample that fairly represents the color space distribution of the galaxies in the imaging survey. Moreover, even if such a representative sample is obtained in a particular patch of sky, sample variance due to the relatively narrow sky coverage of current and near-future calibration samples has a significant impact on the accuracy of the calibration \cite{cunha_etal11}.  We sought to provide general guidelines for a broad range of future imaging surveys, and so we have not attempted to model how these important, survey-specific issues affect the calibration requirements. Instead, in our formulation the photo-z precision requirements are formally specified in terms of the necessary amount of prior knowledge on the photo-z distribution, which in turn is encoded by the parameter $\nspec.$ 

Our modeling of galaxy clustering has several simplifying assumptions that are relevant to the calibration of future imaging surveys. In modeling galaxy bias as a function of redshift only, we have implicitly assumed perfect knowledge of how galaxy bias depends on wavenumber. Uncertainty in the scale-dependence of the galaxy bias is degenerate with uncertainty in $\pmatk,$ and so the improvement in the constraining power of a survey provided by including galaxy clustering will degrade when accounting for uncertainty in the scale-dependence of galaxy bias. However, we have restricted the range of angular scales on which galaxy correlations are exploited so that we probe only very large scales ($\sim50\hinvmpc$), and so we expect corrections accounting for this scale-dependence are small.  

The benefit of galaxy correlations to the photo-z calibration is diluted by lensing magnification bias \cite{bernstein_huterer09, schmidt_etal09}, which induces a spurious correlation between sources that are well separated in redshift space. This effect thus threatens the ability of galaxy cross-correlations to detect and calibrate outliers in the photo-z distribution and will need to be accounted for in order to fully realize the potential of galaxy clustering. We intend to generalize our results to include these effects in a future paper.

\section{Conclusions}
\label{section:conclusions}

We have studied the matter power spectrum calibration requirements for future very-wide-area weak lensing surveys such as LSST or Euclid. While our findings apply to all planned imaging surveys designed to use weak lensing to study dark energy, we have phrased our conclusions in terms of these particular surveys because their calibration demands are the most stringent.  Our results generalize previous findings by simultaneously accounting for photometric redshift uncertainty, as well as by studying the significance of galaxy clustering information. We explored two different models for uncertainty in the nonlinear physics of gravitational collapse, which we describe in detail in~\S\ref{subsection:nlcompare}. In our first model, we assume that the Halo Model accurately predicts the gross shape of $\pmatk$, but that the internal structure of halos is uncertain. In our second, more agnostic model, we allow the value of $\pmatk$ to vary freely about its fiducial value in ten logarithmically-spaced bins of bandpower spanning the range $0.01$ $\hmpcinv\le k \le 10$ $\hmpcinv.$  We conclude this manuscript by providing a brief summary of our primary results.

\ben
\item Future imaging surveys will be unable to self-calibrate the value of $\pmatk$ to better than $7\%$ on any scale. This renders infeasible the possibility of completely self-calibrating the theoretical prediction for the matter power spectrum because systematic errors at such levels would induce unacceptably large biases in the inferred value of the dark energy equation of state. Moreover, the marginalized constraints on $\pmatk$ are the tightest at scales $k\approx0.2$ $\hmpcinv,$ nearly an order of magnitude larger in size than where the unmarginalized constraints computed in Ref.~\cite{huterer_takada05} attain their minimum, emphasizing the necessity of a precise calibration of $\pmatk$ over the full range of wavenumbers $0.1$ $\hmpcinv\lesssim k\lesssim 5$ $\hmpcinv.$
\item To ensure that systematics are kept at levels comparable to or below the statistical constraints on $\wzero$ and $\wa,$ $\pmatk$ must be accurately predicted to a precision of $0.5\%$ or better on all scales $k\lesssim5$ $\hmpcinv$ in advance of future weak lensing observations that will be made by LSST or Euclid.
\item The required precision for the $\pmatk$ prediction as well as the scale to which this precision must be attained depend sensitively on $\ellmax,$ the maximum multipole used in the cosmic shear analysis. Figures~\ref{fig:prec_v_ellmax} and~\ref{fig:kreq_v_ellmax} together provide a concrete guideline that can be used to directly inform the optimal choice for $\ellmax$ that balances the need for statistical precision against the threat of matter power spectrum systematics.
\item In keeping with the results in the Appendix of Ref.~\cite{hearin_etal10}, we find that the photo-z calibration requirements are less stringent by a factor of $\sim3$ when the nonlinear evolution of $\pmatk$ is modeled with the Smith et al.~fitting formula relative to Peacock \& Dodds, significantly relaxing the demands for photo-z precision that appear in Ref.~\cite{ma_etal06}.
\item Dark energy constraints degrade $\sim40\%$ more slowly with photo-z uncertainty when including galaxy correlations in a joint analysis with weak lensing, even when the clustering information is restricted to degree-scales and with coarse tomographic redshift binning so that Baryonic Acoustic Oscillation features are not resolved. 
\item Including galaxy clustering statistics (again, even when Baryon Acoustic Oscillation information is neglected) also significantly relaxes the calibration requirements and mitigates the severity of systematic errors induced by erroneous predictions for $\pmatk,$ especially when prior information on the photo-z distribution is weak. Dark energy systematics can be reduced by up to $50\%$ by including galaxy clustering information; the statistical constraints on $\wzero$ and $\wa$ can degrade $2-5$ times more rapidly with $\pmatk$ uncertainty when galaxy correlations are neglected.
\item The redshift-zero, mean halo concentration, $c_0,$ must be accurately predicted with a precision of $5\%$ or better to keep systematics in dark energy parameters below the level of statistical constraints. If internal halo structure is the dominant mode of $\pmatk$ uncertainty, then the prospect for self-calibrating $c_0$ are quite promising, as this would only degrade the dark energy constraints by $5-10\%.$ 
\item The matter power spectrum calibration requirements are more stringent when the distribution of photometric redshifts is known with less precision. This effect is due to a degeneracy between source redshift and lens size, and is the chief motivation for a simultaneous account of these sources of uncertainty. We find that the constraints on $\wzero$ and $\wa$ degrade $2-3$ times more rapidly with $\pmatk$ uncertainty for spectroscopic calibration samples with the statistical equivalent of $\nspec\approx5000$ relative to $\nspec\approx10^5.$
\item The requirements for the precision with which $\pmatk$ need be predicted are, in general, less stringent for DES than for LSST by a factor of a few. To ensure that matter power spectrum systematics do not contribute significantly to the dark energy error budget for DES, we find that if correlations up to a maximum multipole of $\ellmax=3000$ are used in the lensing analysis then $\pmatk$ will need to be calibrated to an accuracy of $2\%$ or better on scales $k\lesssim5$ $\hmpcinv.$ The DES requirements scale with $\ellmax$ in a similar fashion to the scaling of the LSST requirements summarized in Fig.~\ref{fig:kreq_v_ellmax}. The difference between the requirements is driven by the relative depth of these two surveys: the constraining power on dark energy provided by cosmic shear measurements derives chiefly from small-scale correlations ($k\gtrsim1\hmpcinv$) where shot noise is most significant. The shallower depth of DES ($z_{\mathrm{med}}=0.7,$ $\na=15$ $\mathrm{gal/arcmin^2}$) results in these modes being less informative, and so DES suffers less from uncertainty in small-scale information.
\een

%---------------------------------------------------------------------------------------

\acknowledgments

We thank Wayne Hu, Martin White, Dragan Huterer, Jeff Newman, Chris Purcell, Douglas Rudd, Shahab Joudaki, 
and Bob Sakamano for useful discussions.  ARZ and APH are supported by the Pittsburgh 
Particle physics, Astrophysics, and Cosmology Center (PITTPACC) at the University of Pittsburgh, 
and by the US National Science Foundation through grant AST 0806367. 
ZM is supported through Department of Energy grant DOE-DE-AC02-98CH10886.

\bibliography{wlnlpz}

\end{document}